\documentclass[letterpaper, 10 pt, conference]{ieeeconf}  

\IEEEoverridecommandlockouts                              
\usepackage{graphicx} %
\usepackage[super]{nth}
\usepackage{amsmath} 

\newcommand\norm[1]{\left\lVert#1\right\rVert}
\usepackage{graphicx} %
\usepackage[super]{nth}

\usepackage{etoolbox}  

\patchcmd{\thesection}{\clubpenalty4000}{\clubpenalty10000}{}{}     
\patchcmd{\thesection}{\widowpenalty4000}{\widowpenalty10000}{}{}   

\patchcmd{\thebibliography}{\clubpenalty4000}{\clubpenalty10000}{}{}     
\patchcmd{\thebibliography}{\widowpenalty4000}{\widowpenalty10000}{}{}   
\patchcmd{\bibsetup}{\interlinepenalty=5000}{\interlinepenalty=10000}{}{} 

\title{\LARGE \bf
Catalyzing Clinical Diagnostic Pipelines Through Volumetric\\ 
Medical Image Segmentation Using Deep Neural Networks:\\ 
Past, Present, \& Future
}

\author{ \parbox{5 in}{\centering Teofilo E. Zosa 
        \thanks{In fulfillment of the UCSD Computer Science and Engineering Doctoral Research Mastery Exam}\\
        Department of Computer Science and Engineering\\
        University of California, San Diego\\
        {\tt\small tzosa@ucsd.edu}}
        \hspace*{ 0.5 in}
}

\begin{document}


\maketitle
\thispagestyle{empty}
\pagestyle{empty}

\begin{abstract}

Deep learning has made a remarkable impact in the field of natural image processing over the past decade. Consequently, there is a great deal of interest in replicating this success across unsolved tasks in related domains, such as medical image analysis. Core to medical image analysis is the task of semantic segmentation which enables various clinical workflows. Due to the challenges inherent in manual segmentation, many decades of research have been devoted to discovering extensible, automated, expert-level segmentation techniques. Given the groundbreaking performance demonstrated by recent neural network-based techniques, deep learning seems poised to achieve what classic methods have historically been unable. 

This paper will briefly overview some of the state-of-the-art (SoTA) neural network-based segmentation algorithms with a particular emphasis on the most recent architectures, comparing and contrasting the contributions and characteristics of each network topology. Using ultrasonography as a motivating example, it will also demonstrate important clinical implications of effective deep learning-based solutions, articulate challenges unique to the modality, and discuss novel approaches developed in response to those challenges, concluding with the proposal of future directions in the field.

Given the generally observed ephemerality of the best deep learning approaches (i.e., the extremely quick succession of the SoTA), the main contributions of the paper are its contextualization of modern deep learning architectures with historical background and the elucidation of the current trajectory of volumetric medical image segmentation research.

\end{abstract}

\section{INTRODUCTION}

Medical imaging is a key aspect of modern medical care. It enables the detection, identification, and monitoring of lesions, tumors, and other abnormalities; a necessary prerequisite for medical diagnostics, planning, and guidance related to many pathologies. Clinically significant radiology modalities include computed tomography (CT), magnetic resonance (MR), functional magnetic resonance (fMR), positron emission tomography (PET), X-ray radiography (X-ray), and ultrasonography, among others.

A central component of many modern medical treatments is segmentation of these images into regions of interest to inform clinical diagnostics, either directly or in conjunction with other methods in the diagnostic pipeline (e.g. registration); assist in surgical planning; or even provide real-time feedback for physically invasive procedures. Unfortunately, manual segmentation is a difficult, time-consuming, error-prone process that is often the bottleneck in critical clinical workflows due to significant intra- and inter-rater variability. Because of this, much research has been conducted on computer-assisted automatic segmentation to assist clinicians in providing more effective, efficient, and affordable care. 

\begin{figure}[ht] 
    \centering
    \includegraphics[clip=true, trim=0pt 0pt 0pt 0pt, width=0.4\textwidth]{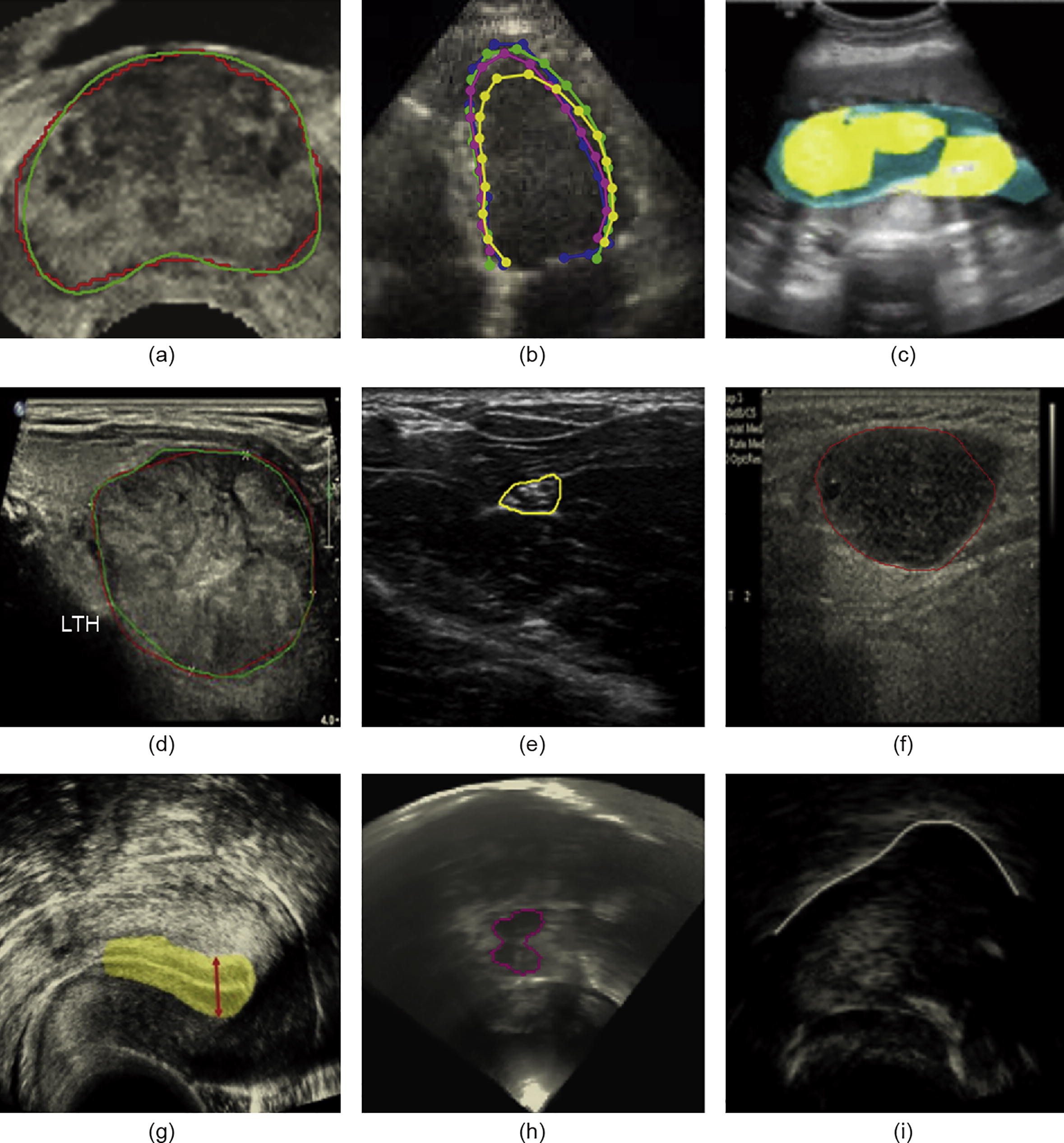}
    \caption{``Examples of segmentation results from certain anatomical structures using deep learning. (a) prostate \cite{Yang2016}; (b) left ventricle of the heart \cite{Nascimento2016}; (c) amniotic fluid and fetal body \cite{Li2017}; (d) thyroid nodule \cite{Ma2017}; (e) median nerve structure \cite{Hafiane2017}; (f) lymph node \cite{Zhang2017}; (g) endometrium \cite{Singhal2017}; (h) midbrain \cite{Milletari2017}; (i) tongue contour \cite{Aslan2018}. All of these results demonstrated a segmentation performance that was comparable with that of human radiologists. Lines or masks of different colors represent the corresponding segmented contours or regions'' \cite{Liu2019}}
    \label{fig:segmentation_ex}
    \vspace{-5mm}
\end{figure}

Formally, semantic segmentation of images is defined as the identification of the set of pixels (or voxels in the case of 3D volumetric images) that make up the object of interest. This typically comes in the form of a segmentation mask or contour outline aligned to the target anatomy (see Figure \ref{fig:segmentation_ex} \cite{Liu2019}). In traditional medical image segmentation, a trained operator, usually a physician or radiologist, annotates medical images to delineate objects of interest. These are usually either 2D images or 2D slices of a 3D volume sliced along one of three coordinate axes. 

Computer-assisted automatic segmentation began in the 1970s with lower-level methods such as thresholding, edge detection, and region growing. These were followed by higher-level methods such as atlases, deformable models, and statistical classifiers. Finally, a type of statistical classifier known as an artificial neural network (ANN) rose to prominence, with a subclass of ANN, the convolutional neural network (CNN) becoming the core of many current SoTA segmentation algorithms. While neural network-based approaches generally lead to stellar performance, they have yet to achieve parity with expert annotators. Additionally, their applications are skewed towards relatively easier-to-segment imaging domains, such as CT and MRI, which produce high contrast, high resolution images with consistent spatial parameters, sometimes in multiple co-registered modalities. Thus, much work remains to be done both in improving the performance of these methods as well as extending them to other less researched difficult-to-segment medical imaging domains such as ultrasound. 

This paper will continue with a brief introduction to each of the aforementioned segmentation methods in Section \ref{background}. Section \ref{networks_overview} will then give a brief overview of CNNs, Section \ref{networks} will continue discussing CNNs in the context of volumetric medical image segmentation, and compare and contrast some of the more recent SoTA methods across different imaging modalities and benchmarks, highlighting their particular contributions and identifying the current trend towards greater integration of contextual information. Section \ref{ultrasound} will discuss deep learning in the context of ultrasound specifically, outlining the unique challenges and potential impact of fully-automated expert-level solutions as well as corresponding contemporary approaches. Section \ref{future} will then briefly propose future directions followed by the paper's conclusion in Section \ref{conclusion}.

\section{Background}
\label{background}

In the domain of automated medical image analysis, a few common features include image element (pixel/voxel) intensity, gradient magnitudes, and texture measures \cite[pg. 73]{Royan2012}. Broadly speaking, segmentation algorithms can be grouped into one of three categories: classification-based, which assigns class labels to individual image elements, computing the segmentation mask directly; edge-based, which uses inter-region similarity to compute segmentation maps; and region-based, which uses intra-region similarity to compute segmentation maps \cite[pg. 73]{Royan2012}. Segmentation algorithms can also be characterized by the type of information explicitly considered. Low-level methods compute segmentation masks or boundaries directly from image elements, generally without consideration of the broader context from which the images are derived. In contrast, high-level methods make use of information such as the shape or texture of the target segmentation object, using this information to fit sophisticated mathematical models to the data to yield segmentation results. 

\subsection{Low-Level Methods}
The earliest segmentation algorithms worked directly on low-level features; in other words, strictly on information derived from the intensity values of the image elements. They are inherently unable to take advantage of contextual information which cannot be encoded by element intensity and are generally local methods, only considering image elements in a sub-region of the image. Consequently, they are most effective when objects of interest have a high contrast in relation to the rest of the image. Common low-level methods include thresholding, edge detection, and region growing.

\subsubsection{Thresholding}
Thresholding is a classification-based technique, performing point-based binary classification on individual image elements, with classification based on a comparison between the element's intensity value and a threshold value. In the case of multi-region classification, multiple thresholds can be built into the threshold criteria. Thresholding can be global (i.e., a constant threshold throughout the image) or local (i.e., spatially varying with the image). In contrast to point-based thresholding which considers single image elements in isolation, region-based thresholding considers neighboring image elements jointly. Threshold values may be known a priori or estimated from intensity histograms using shape-based estimation (histogram features) or optimal estimation (optimization of an objective function) \cite[pg. 78]{Royan2012}.

The main strength of this method is that in certain conditions it can be quite effective relative to its simplicity. Its main drawback is that it is very brittle, generally producing unacceptable performance in the majority of non-trivial cases. 

\subsubsection{Edge Detection}
A widely observed phenomenon in images is the presence of edges, or adjacent points with significant differences in intensity values, which oftentimes delineate disparate object. If the assumption can be made that detectable edges exist along the entirety of target objects' true boundaries, then edge detection algorithms can be leveraged.

Edge detection algorithms are edge-based methods which seek to compute boundaries by detecting edges between objects of interest under the assumption that boundaries have corresponding edges, segmenting images based on these boundaries. Edge detection algorithms generally work by computing first- or second-order derivatives to identify these points. These can be efficiently computed via a convolution operator over image regions, with Sobel, Prewitt, Canny, and the Laplacian-of-Gaussian operators being among the most historically popular convolutional kernels in edge detection \cite[pg. 88]{Royan2012}. The morphological gradient, computed via a series of erosion and dilation operations defined by the field of mathematical morphology, and related techniques are also popular and tend to be very useful \cite[pg. 177-261]{Royan2012}. 

While effective for some tasks, edge detection is not without its faults. Namely the assumption of entire boundaries consisting of detectable edges, a violation of which will result in incomplete segmentation due to the presence of discontinuous edges, and the choice of specific edge-detection algorithm being dependent on the target domain and task. In addition, classic edge detection algorithms are not robust to image variations such as noise, artifacts, or lighting. This leads to the need for hand-tuned pre- and post-processing steps to ameliorate the complications these variations induce \cite[pg. 140-180]{JainRameshandKasturiRangacharandSchunck1995}.

\subsubsection{Region Growing}
In contrast to edge detection, region-growing is a region-based method that seeks to explicitly define the regions objects of interest inhabit as opposed to implicitly defining objects as anterior or posterior to a boundary. 

Region-growing takes as input specific points inside of an image, known as seed points, which are initialized as singleton sets. For each element in each set, the algorithm compares the element to adjacent pixels that belong to another set. If the algorithm identifies them as belonging to the same region based on similarity metrics, it merges the two sets together. The algorithm terminates when there are no remaining mergeable sets \cite[pg. 98]{Royan2012}.

The main drawbacks of region growing are similar to those of edge detection, not being particularly robust to image variations (i.e., noise, artifacts, lighting, etc.). Consequently the choice of seed points matter and sometimes must be set manually \cite[pg. 203]{Norouzi2014}. Additionally, similarity metrics must be well-defined as false-positives can cause regions to occupy a much larger area than the target object, potentially consisting of the entirety of the image (akin to edge discontinuities in edge detection algorithms). Finally, it is relatively more computationally expensive than thresholding or edge detection when employing automatic seed point initialization algorithms, especially in the presence of noisy images \cite{Adams1994}.

\subsection{High-Level Methods}
High-level image segmentation methods were developed in response to the lack of robustness to image quality, object shapes, noise, and sampling artifacts displayed by lower-level methods which consequently either fail entirely or require extensive additional pre- and post-processing \cite[pg. 131]{Royan2012}. These methods utilize higher-level features and mathematical abstractions to counter the aforementioned considerations. Examples of popular methods include atlas-based techniques, deformable models, and deep neural networks. 

\subsubsection{Atlas-based}
The fact that many anatomical objects vary very little in shape and structure across patients enables a method known as atlas-based segmentation. Atlas-based techniques leverage a database of labeled objects of interest, known as an atlas, and attempt to fit the labeled object to the image data via a set of coordinate transforms.  The best set of transforms is determined by a similarity metric, oftentimes pixel or voxel-wise intensity. The series of transforms usually involves a rigid transformation stage to align the anatomies, followed by a non-rigid transformation stage to fine-tune the model to the target anatomy.

The performance of atlas-based methods relies on the constraints placed on the transformations, namely non-rigid transformation. These constraints must allow enough expressiveness to sufficiently handle complex anatomy while simultaneously adhering to the true shape of the target object, requiring a model that is consistent with realistic deformations of related materials (i.e., elastic solids and viscous fluids) \cite[pg. 429-430]{Royan2012}; a non-trivial matter as these are themselves ongoing areas of research.  

\subsubsection{Deformable Models}
Deformable models enable an edge-based segmentation technique wherein models can be deformed and moved toward the object boundary to achieve boundary detection and resultant segmentation.

Deformable models, popularized by Kass, Witkin, and Terzopolous \cite{Kass1988} with their active contour formulation also known as `snakes', are curves, or surfaces in the case of 3D volumetric images, in an image that move in relation to internal and external forces relative to the curve or surface. These forces jointly fit the model to the object of interest, with external forces moving the model toward the object and internal forces maintaining smoothness of the model during deformation. Internal forces are dictated by the particular deformable model employed, while external forces are computed from the image data. When used in conjunction with a priori shape information, the model smoothness constraint affords deformable models a robustness to noise and boundary gaps, a problem common in classic edge-detection-based segmentation techniques. Furthermore, it enables a consistent and rigorous mathematical description of object boundary elements that provides interpretability and can be used by other downstream applications for other purposes \cite[pg. 131]{Royan2012}.

Deformable models can be categorized into two types: parametric and geometric. Parametric deformable models explicitly encode curves or surfaces in their parametric forms during deformation while geometric deformable models represents curves or surfaces implicitly as a level set \cite{Osher1988} of a higher dimensional function \cite[pg. 133]{Royan2012}.

Parametric models allow direct interaction with the model and can lead to mathematically compact representations suitable for real-time applications \cite[pg. 133]{Royan2012}. However, parametric models suffer from two major limitations. First, if the model is initialized with a shape and size dissonant with the target object, the model must be reparameterized dynamically. This may be computationally expensive, adversely affecting the feasibility of this approach. Second, in the case of multiple objects or objects with unknown topology, parametric models are sometimes unable to cope with the need for certain topological adaptations (i.e., splitting or merging model components) \cite[pg. 145-146]{Royan2012}.

Geometric models were proposed to address the limitations of traditional parametric models. In this paradigm, curves or surfaces are evolved geometrically with parameterization being computed after complete model deformation, making model-fitting independent of the parameterization and enabling automatic topological adaptations, being especially useful when the topology is not known in advance \cite[pg. 146]{Royan2012}. However, this topological adaptivity is also a weakness as the model may conform to specious shapes due to the presence of noise or incomplete boundary information \cite[pg. 154]{Royan2012}.

It should be noted that in both cases these models are often enhanced by the inclusion of a priori information (e.g. shape constraints) and global shape properties (e.g. orientation and size constraints) \cite[pg. 154-155]{Royan2012}.

\subsubsection{Deep Neural Networks}
Artificial feed-forward neural networks, more commonly referred to as simply ``neural networks'' are at the core of many modern image processing techniques. Neural networks are universal function approximators originating from the field of machine learning, achieving function approximation via a series of mathematical operations whose parameters are automatically learned via an iterative training process. 
The structure of a neural network enables hierarchical feature extraction, where each computation in the cascade automatically extracts salient features from the input feature space via the learned parameters at that step which, when properly optimized, come to represent task-relevant relationships between incoming features. 


When applied to the task of medical image segmentation, a special type of neural network known as a convolutional neural network (CNN) is typically employed and segmentation is generally framed as a classification problem over each image element. 

CNNs naturally encapsulate or extend many of the classical segmentation methods, exploiting their strengths while overcoming many of their limitations. For instance, thresholding can be implicitly performed by the neural net in its mathematical operations to encode feature saliency.  In addition, when applied to images, the hierarchical feature extraction inherent in CNNs tends to identify image primitives such as edges, implicitly performing edge detection.  Further downstream this becomes shape detection and ideally culminates in object detection, affording CNNs a knowledge representation that includes high-level descriptions of complex anatomical properties. 

A strength of CNNs is mutual feature detection and a varying level of translation invariance afforded by the way CNNs generate latent feature spaces \cite{Bengio2013}. In combination with the increasing ubiquity of sufficiently large datasets in different domains, CNN-based methods have yielded groundbreaking performance on many tasks across many fields, especially in the domain of medical imaging \cite{Litjens2017, Liu2019}. 

When hand-tuning classical methods to detect objects of interest, many of these methods were not robust to object variability and performed poorly in non-ideal cases. On the contrary, when a neural network with sufficient representational power is given enough input data containing a rich variety of examples for each class, it can automatically extract more general class-relevant features to produce more robust classification, thus solving the recognition problem for non-ideal class variants in a very extensible manner. Additionally, in classification tasks, neural networks often output probability values which can be used to indicate confidence, enabling retrieval of an ordered list of most probable classes. Neural networks can also update their knowledge base in the face of new data without the need to train a new model from scratch. 

Another advantage comes from the opportunity for \textit{transfer learning}, or using the parameters of a neural network trained on a related to task to initialize a new neural network aimed at the specific task at hand. This is especially useful in the domain of medical imaging where there is often a paucity of data, especially of rare pathologies whose variations we are most interested in detecting.

\section{Neural Network Overview}
\label{networks_overview}


Neural networks model the relationship mapping between input data and target outputs mathematically as a series of nested functions. Neural networks commonly abstract fundamental operations into a logical unit known as a node which contains a parameterized linear operation followed by a non-linear \textit{activation function}. The set of nodes in the same level of the hierarchy are said to belong to the same logical module commonly known as a \textit{layer}, and each layer taking as input the feature maps from the previous layer. The layer that operates directly on input data is termed the \textit{input layer}, the layer that produces the final output values is termed the \textit{output layer}, and any layer that exists between the input and output layers is referred to as a \textit{hidden layer}. Each hidden layer can have an arbitrary number of nodes which are also called \textit{hidden units}. 

The parameters of a node's linear operation consist of learnable weights $w$, with the same cardinality as the set of input features, and a bias term $b$. The computations of a node given input features $x$ take the form $f(\sum_iw_ix_i + b))$  where $\sum_iw_ix_i $ is a matrix multiplication of $w$ with $x$. $f$ is commonly the sigmoid, tanh, or ReLU \cite{He2015d} function. In multi-class classification problems, the output layer typically uses a softmax \cite[pg. 64]{Lange1997} activation function to generate a probability map over the output classes. 

Neural networks that use multiple hidden layers to exploit hierarchical feature extraction are termed deep neural networks (DNNs). DNNs are able to take low-level features in the form of raw input data and automatically build increasingly higher-level feature representations as the depth of the network (number of layers) increases. This enables robust automatic feature extraction, obviating the need for feature engineering and pre-processing that is oftentimes brittle, resource-intensive, and application-specific \cite{Bengio2013}.

\subsection{Factors Affecting Performance}
In neural networks, function approximation fidelity is determined by the network's representational power as well as the quality and quantity of input data.

Representational power, or how well the neural network can approximate an arbitrary function, is dictated by parameters of the neural network known as \textit{hyperparameters} that are set prior to training. These include the choice of activation functions, optimizer, learning rate, regularization scheme, number of layers, number of units per layer, weight initialization scheme, etc. 

Given the nature of neural networks, the main route to increasing representational power is by deepening (adding more layers) or widening (adding more hidden units) the network.
The titration of representational power with predictive performance on unseen data is a major consideration. While sufficient representational power is a necessity, a network that is too readily able to discover patterns in the data is prone to a phenomenon known as \textit{overfitting}. Overfitting refers to the fact that the learned patterns are idiosyncratic to the data on which the network was trained and do not generalize well to the population from which the data was sampled, causing a decrease in performance when applied to new data. 

A sufficiently large quantity of high quality training data also helps improve performance and prevent overfitting as it now better approximates the broader population. Though, in practice this may be difficult to achieve, especially in the domain of medical imaging, and comes at the cost of increased training time.

\subsection{CNNs}
Traditional DNNs employ fully-connected layers where each feature (i.e., the output of each node) in a layer is fed in as input to every node in the following layer. If an entire image or volume is fed into a fully-connected neural network, each pixel or voxel becomes an input feature, resulting in an explosion of parameters, even for modest sized inputs. 

In contrast to DNNs, CNNs eschew fully-connected layers in favor of the eponymous \textit{convolutional layer} which exploits inter-domain and inter-image correlations for robust and efficient feature extraction. Within a convolutional layer, convolution is the linear operation and the learnable parameters at that layer are a set of convolutional kernels whose size (also known as the \textit{receptive field size}) is a hyperparameter and whose weights are shared across different regions along the spatial dimension of the input feature space. This improves performance by both reducing the number of learnable parameters and enabling full-image feature detection with a single kernel. 

CNNs have also historically made use of \textit{pooling layers} which apply a permutation-invariant (i.e., max or mean in the case of max pooling or mean pooling, respectively) operation to combine or \textit{pool} pixel values in a region. An important feature is that, similar to convolutional layers, pooling layers yield a varying amount of translation invariance. In conjunction with data augmentation techniques (i.e., scaling, shearing, and rotation), CNNs become particularly robust to variations in rigid deformations of target objects \cite{Bengio2013}. 

\subsection{Neural Network Training}
In the classic supervised learning paradigm, neural networks achieve function approximation by iteratively tuning their learnable parameters in a learning phase known as \textit{training}. This is accomplished through the optimization of a target function defined between the output prediction $y$ given an input example $x$, and known ground truth label $y'$. Typically, the target function is a loss function and common optimization methods, also known as \textit{optimizers}, include stochastic gradient descent \cite{Bottou2010}, ADAM \cite{Kingma2014}, and Laplacian-smoothed gradient descent \cite{Osher2018}, among many others. At each iteration, an optimizer updates each layer's weights based on its contribution to the final prediction $y$ as determined by an efficient gradient computation method known as backpropagation \cite{LeCun1998}.

In segmentation, $x$ typically consists of 2D or 3D images with one or more channels, and $y'$ is an annotated segmentation map with voxel-wise labels. In the case of co-registered multimodal data, modalities can be concatenated together in the channel dimension to provide CNN kernels the ability to integrate inter-modal information simultaneously along the same spatial dimension, automatically finding the high-level relationships amongst these modalities.

\subsection{Modern Modifications}
Given empirical performance observations, modern CNN-based image analysis methods typically use ReLU (or some variant) as the non-linear activation function and ADAM as the specific gradient descent optimization algorithm. Other significant modifications include:

\subsubsection{Removing Pooling}
Depending on the task, it is also common to eschew pooling layers and replace them instead with convolutional layers. This has been shown to improve the performance of modern neural networks in certain tasks and reduce the memory footprint \cite{Springenberg2014}.

\subsubsection{Fully Convolutional Neural Networks}
To address computational redundancy and flexibility of previous CNN formulations in dense semantic segmentation tasks, fully-convolutional neural networks (fCNNs) have been proposed whereby  fully-connected layers are efficiently rewritten as convolutions, allowing the network to output a probability map over each pixel/voxel rather than over each class. This modification enables variable-sized inputs making it more generalizable in addition to more computationally efficient \cite{Long2015}. 

\subsubsection{Residual Learning}
While deeper networks should always meet or exceed the performance of shallower networks in theory, in practice simply increasing network depth introduces a paradoxical degradation of performance after a certain point. 

\begin{figure}[ht] 
    \centering
    \includegraphics[clip=true, trim=0pt 0pt 0pt 0pt, width=0.4\textwidth]{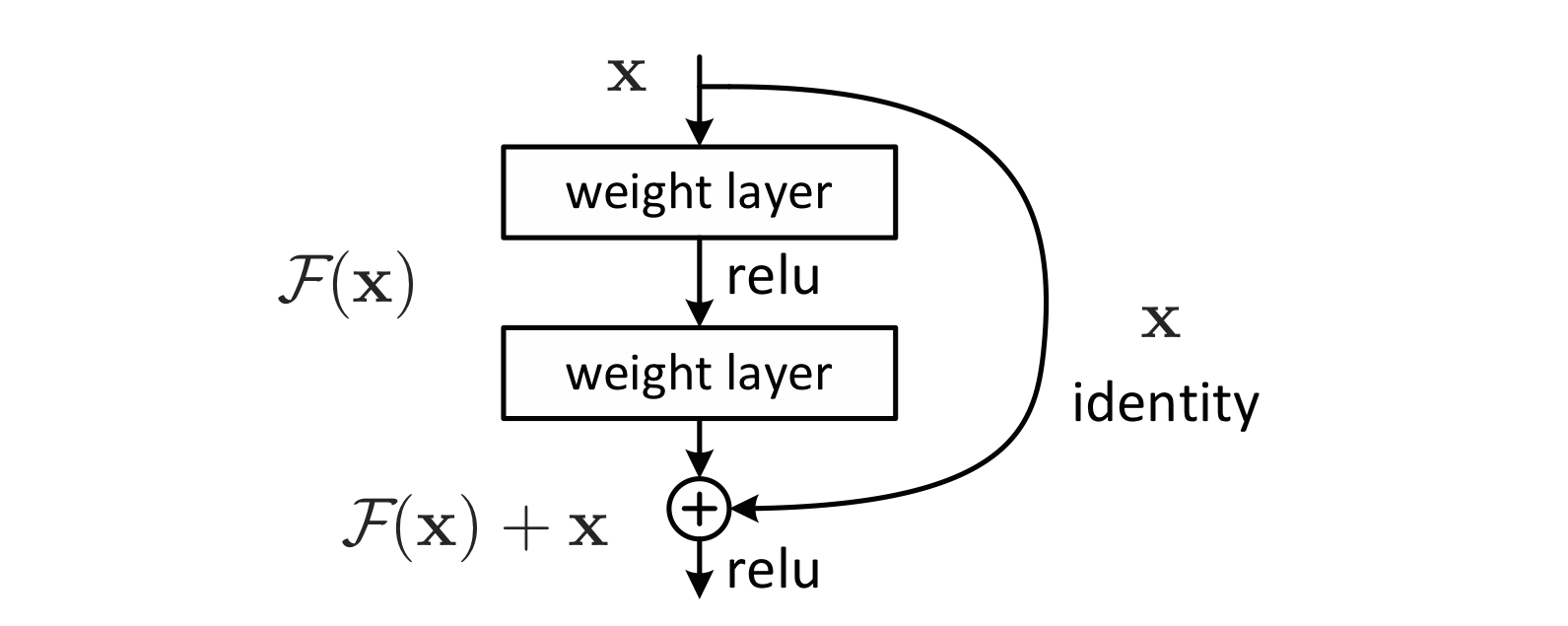}
    \caption{Residual block illustration \cite{He2016}}
    \label{fig:resblock}
    \vspace{-5mm}
\end{figure}

He at al. \cite{He2016} proposed a solution to this problem by reframing the learning process as explicit layer-wise learning of the residual function between a layer and output from an upstream layer. Concretely, they propose the idea of \textit{residual blocks}, or logical computational units consisting of one or more layers of the neural network which directly learn the residual function of a previous layer (rather than assuming it is learned implicitly) via ``skip connections'' which add the output of a hidden layer's activations with the linear outputs of another layer further downstream (see Figure \ref{fig:resblock}) \cite{He2016}. At a high level, the implication of this topological modification is that the skip connection allows a residual block to simply learn the identity function if no useful higher-level features can be learned at that block, rendering accuracy no worse than if the block simply had been excluded from the network. 

Indeed, the authors found this to be an effective way to stabilize training for very deep networks, leading to improved performance while also obviating the need for strong regularization schemes. This has been a major boon to the deep learning community with most modern neural network architectures leveraging residual learning to some degree.

\subsection{Common Evaluation Metrics}
Finally, common metrics used for evaluating segmentation predictions against ground truth data \cite{Taha2015}:

\subsubsection{\textbf{Dice Similarity Coefficient (DICE)}}
A measure of similarity between two segmentations $X$ and $Y$ of the form:
    \begin{equation*}
        DICE(X, Y) =\frac{2\times|X\cap Y|}{|X| + |Y|}
    \end{equation*}
where $X$ and $Y$ are sets of image coordinates. The Dice score ranges from $[0-1]$ with higher scores indicating a more accurate predicted segmentation (i.e., if $X=Y$ then the Dice coefficient is equal to 1). By far the most widely used metric.

\subsubsection{\textbf{Hausdorff Distance (HD)}} 
    
The greatest distance given some distance metric between a point in $X$ and the closest point in $Y$, and vice-versa. In other words, 
    \begin{equation*}
        HD (X, Y) = \max ({h(X, Y), h(Y, X)})
    \end{equation*}
where 
    \begin{equation*}
    h(X, Y) = \max_{x \in X}\min_{y \in Y} \norm{x-y}
    \end{equation*}
Lower values of HD imply more accurate predicted segmentations. HD is sensitive to outliers, so the average HD (AHD) is also sometimes used where 
    \begin{equation*}
        AHD (X, Y) = \max({h_{avg}(X, Y), h_{avg}(Y, X)})
    \end{equation*}
 and 
    \begin{equation*} 
    h_{avg}(X, Y) = \frac{1}{N} \sum_{x \in X}\min_{y \in Y} \norm{x-y} 
    \end{equation*} 
or the average HD for all points in $X$.


\subsubsection{\textbf{Intersection Over Union (IOU)}} 
    
a measure of the region size segmentations $X$ and $Y$ share in common divided by the region size covered by either, of the form:
    \begin{equation*}
        IOU(X, Y) =\frac{|X\cap Y|}{|X\cup Y|}
    \end{equation*}
 where $X$ and $Y$ are sets of image coordinates. Like DICE, the IOU score ranges from $[0-1]$ with higher scores indicating a more accurate predicted segmentation.

\section{Volumetric Segmentation Networks}
\label{networks}

Recent neural network-based approaches to volumetric segmentation in medical imaging can be roughly divided into those that work in 2D and those that work in 3D. \footnote{While some authors have defined a 2.5D classification space which divides a 3D space into $<$3D subspaces (i.e., slices, 3D patches, etc.), this paper will attempt to segregate them into methods that use 2D operations or 3D operations, making note of these hybrid approaches as needed.}

\subsection{2D vs. 3D}
The main advantage of 2D networks is that they are more computationally tractable, working on images rather than full volumes which are exponentially more complex. Their main drawback is that they do not adequately make use of the surrounding spatial context from which the image was derived. As a result, extra training, inference, and post-processing is oftentimes necessary to achieve sufficient segmentation performance, for example, processing a volume slice-by-slice along all three orthogonal planes requires 3x the computation as compared to a fully 3D approach. 

3D networks overcome these drawbacks by integrating greater amounts of surrounding context around a target object, though doing so at the cost of reduced training and inference speed or representational power due to the dramatic increase in computation and memory requirements. 

\subsection{Integrating Contextual Features}
Another consideration is the balance of global and local features. Global features are peripheral to, and spatially distant from, the target object, providing information on the location of the segmentation target relative to the overall anatomy (in addition to information idiosyncratic to the anatomy itself). Local features directly inform the segmentation by providing information on the appearance of the segmentation target object itself (in addition to the immediate periphery). 

Whether operating in 2D or 3D space, both global and local features of a target object are needed for accurate segmentation. For instance, in white matter hyperintensity (WMH) segmentation, lesions exhibit low spatial continuity and vary in both size and location across patients. This necessitates knowing both the larger context of the specific patient's brain as well as the high-resolution local information in each neighborhood of the brain to determine if certain regions are lesions and, if so, how they should be segmented. This presents a major challenge as classic methods operate on input at a specified scale and do not have an intelligent way of simultaneously integrating multi-scale information. 

Another major complicating factor is that there is no absolute delineation between local information and global information. Instead, contextual information lies on a spectrum with notions of local and global labels lying on polar extrema. Consequently, even if a single scale was optimal, this scale would need to be discovered empirically for each segmentation task.

\subsection{The Evolution of Modern Segmentation Networks}


\subsubsection{\textbf{U-Net}}
\begin{figure}[ht]
  \centering
  \includegraphics[clip=true, trim=0pt 0pt 0pt 0pt, width=0.49\textwidth]{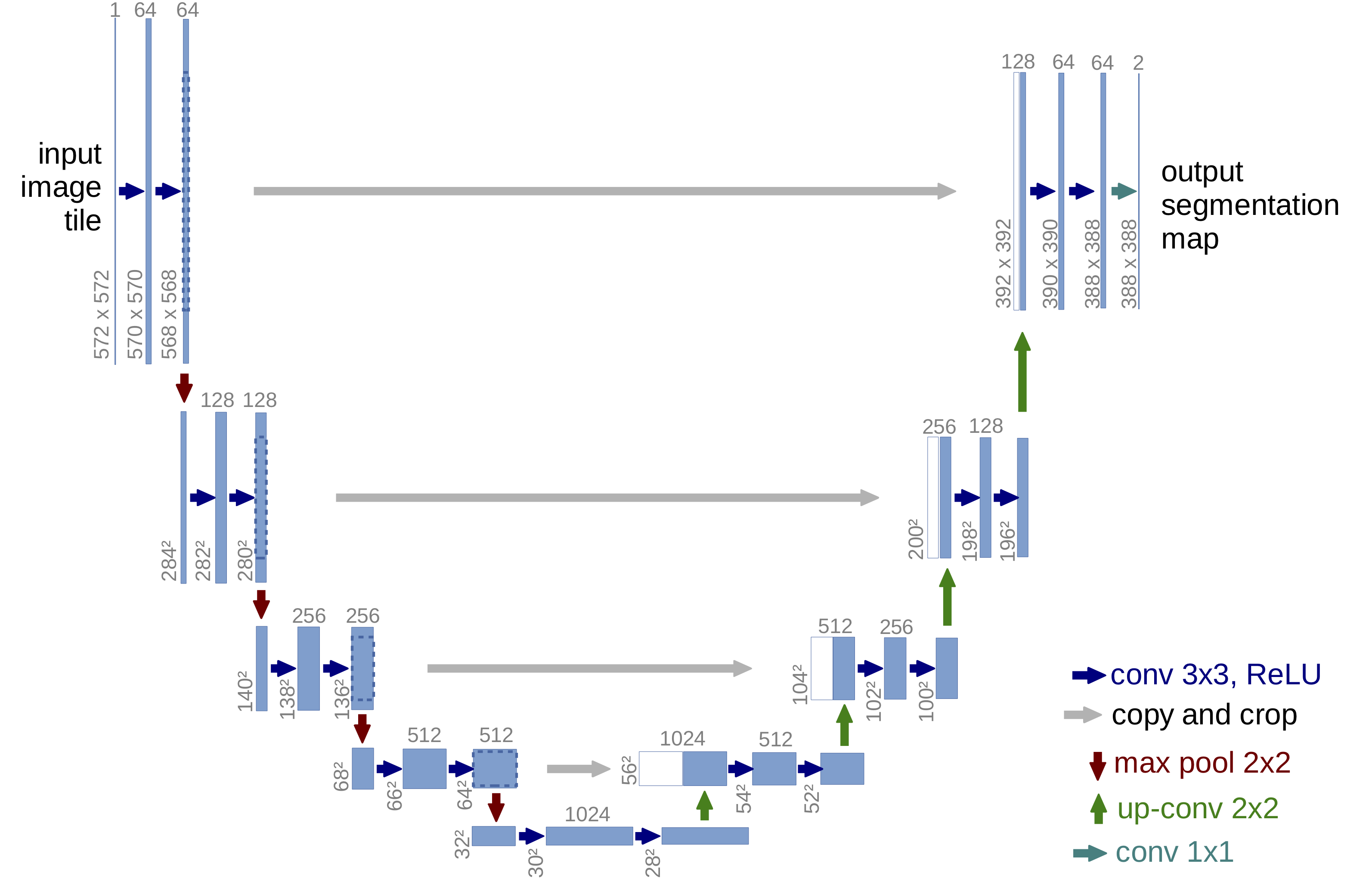}
  \caption{U-Net: ``U-net architecture (example for $32x32$ pixels in the lowest resolution). Each blue box corresponds to a multi-channel feature map. The number of channels is denoted on top of the box. The x-y-size is provided at the lower left edge of the box. White boxes represent copied feature maps. The arrows denote the different operations.'' \cite{Ronneberger2015} 
  }
  \label{fig:u-net}
\end{figure}

Ronneberger et al. \cite{Ronneberger2015} sought to address the issue of both local and global feature integration with their proposal of U-Net, the progenitor to many modern deep learning models applied to segmentation which takes its name from the ``U'' shape the canonical graphical representation of the network resembles (see Figure \ref{fig:u-net}). 

Up until then, a major issue in segmentation methods was balancing the integration of varying levels of spatial context as input to segmentation methods. As segmentation relies on both local and global contextual information, a successful method would necessarily include both types of information.

U-Net solves this problem via a series of downsampling, upsampling, and feature concatenation operations in its architecture, enabling simultaneous multi-scale feature learning and efficient segmentation map generation with a single forward pass of the network (see Figure \ref{fig:u-net} for an overview of the architecture). 

\begin{figure}[t]
  \centering
  \includegraphics[clip=true, trim=0pt 0pt 0pt 0pt, width=0.5\textwidth]{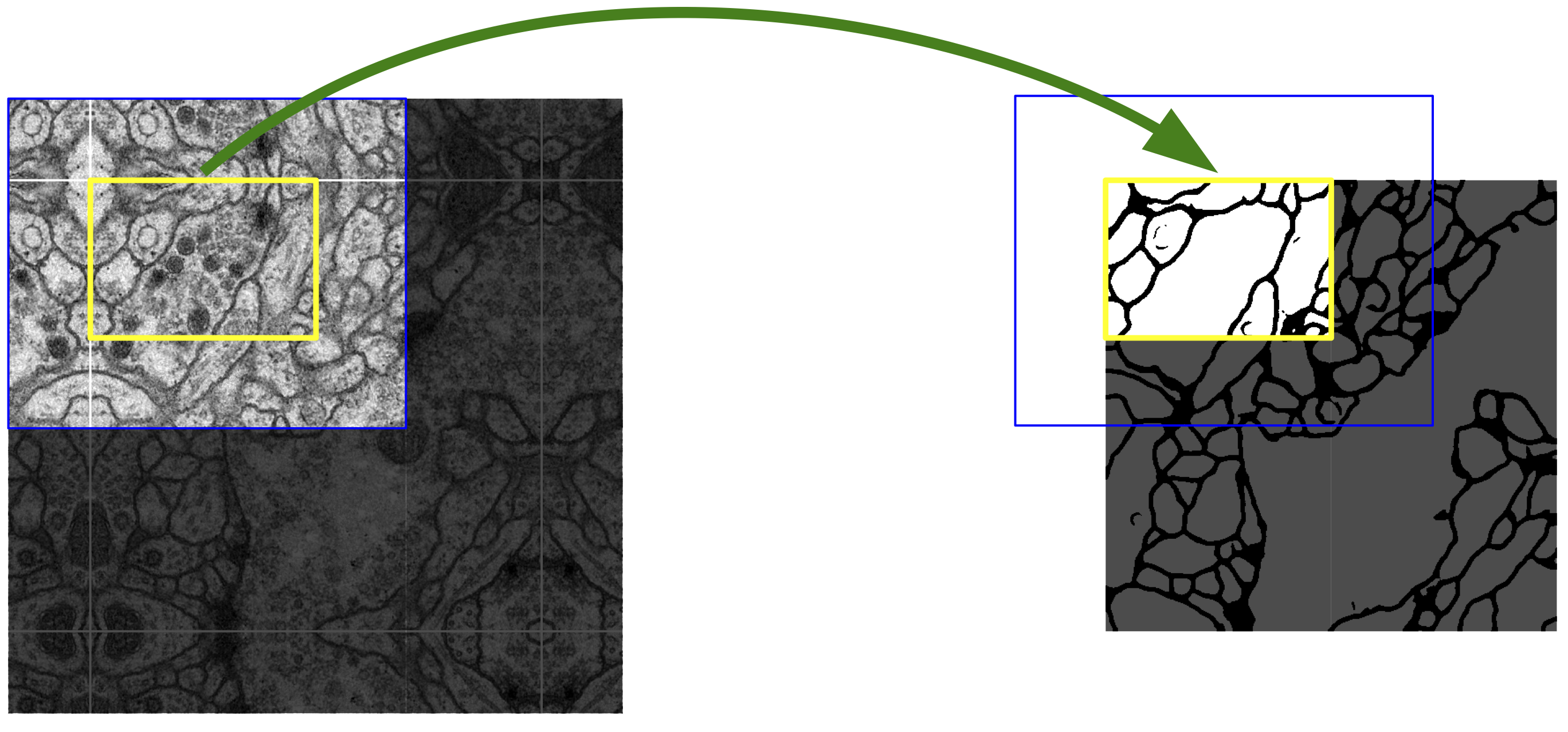}
  \caption{U-Net: ``Overlap-tile strategy for seamless segmentation of arbitrary large images (here segmentation of neuronal structures in EM stacks). Prediction of the segmentation in the yellow area, requires image data within the blue area as input. Missing input data is extrapolated by mirroring'' \cite{Ronneberger2015}}
  \label{fig:unet_overlap-tile}
\end{figure}

The network takes as input $572x572$ grayscale image slices and is divided into two halves: a \textit{contracting path} which comprises the first part of the network and an \textit{expansive path} which comprises the latter half. The contracting half is characterized by the presence of downsampling operations which occur at a rate of $2x$ via max pooling layers with $2x2$ kernels and a stride of 2. The expansive path symmetrically contains all the upsampling layers, with upsampling also occurring at a rate of $2x$ via deconvolution layers with $2x2$ kernels. In the contracting path, the feature maps for each layer immediately preceding a downsampling layer are copied and concatenated to the input features of the corresponding post-upsampling layer in the expansive path after being cropped to matching spatial dimensions. The final layer of the network is a convolutional layer with a $1x1$ kernel which classifies each voxel of the central $388x388$ image crop (see Figure \ref{fig:unet_overlap-tile}). The segmentation task was framed as a pixel classification problem and the authors consequently utilized a cross-entropy loss.


The authors tested U-net on several datasets, notably the ISBI cell tracking challenge 2015 where they bested the previous SoTA on segmentation of brain tumor cells captured by phase-contrast microscopy (9\% improvement in IOU score), and cervical cancer cells captured by differential interference contrast microscopy (31\% improvement in IOU score). 

\subsubsection{\textbf{U-Net with Residual Connections}}
 Drozdzal et al. \cite{Drozdzal2016} explored the use of short and long skip connections in a U-net-like model (modifying \cite{He2016} by adding an expanding path and corresponding connections from the contracting path). They noted that the copy and concatenation of features in U-Net's contracting path with features in the expanding path are akin to long skip connections and so choose to sum rather than concatenate the features in their models. The combination of both short and long skip connections led to better convergence and training stability relative to variants of the network that either utilized only one type of connection, or neither. 

\subsubsection{\textbf{3D U-Net}}
\begin{figure}[ht]
\centering
\includegraphics[clip=true, trim=0pt 0pt 0pt 0pt, width=0.48\textwidth]{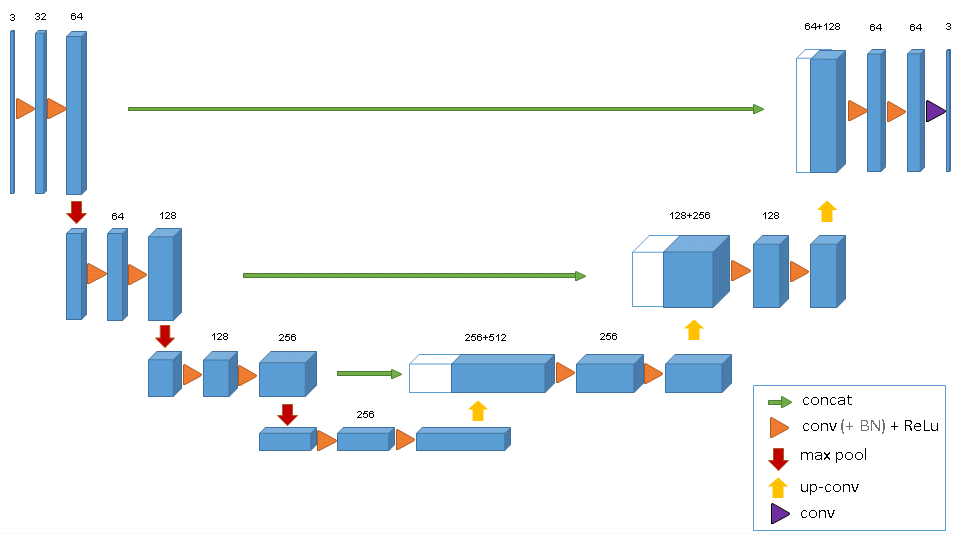}
\caption{3D U-Net: ``The 3D u-net architecture. Blue boxes represent feature maps. The number of channels is denoted above each feature map.'' \cite{Cicek2016}}
\label{fig:unet3d}
\end{figure}

\c{C}i\c{c}ek et al. \cite{Cicek2016} directly extended U-net to process all three spatial dimensions simultaneously, proposing a variant that utilized 3D convolutions in place of 2D convolutions for full volumetric segmentation. Aside from reducing the number of output features in every layer of the the contracting path by half, save for those directly preceding downsampling operations, the 3D U-net was identical to the original U-net (see Figure \ref{fig:unet3d}).

Given sparsely annotated training data (volumes with only a few slices annotated), the authors used the 3D U-net to produce dense volumetric segmentation of Xenopus kidney embryos captured by confocal microscopy in two tasks. The first was a `semi-automated'' segmentation task where dense (complete) volume segmentation was produced for a sparsely annotated training sample, achieving a 7\% higher IOU score relative to a 2D U-Net.

The second was a fully-automated segmentation task where a dense volume segmentation was produced for an unlabeled volume on which the network had not been trained, achieving an 18\% higher IOU compared to a 2D U-Net.

\subsubsection{\textbf{V-Net}}
\begin{figure}[t] 	
\centering 	
\includegraphics[clip=true, trim=0pt 0pt 0pt 0pt, width=0.49\textwidth]{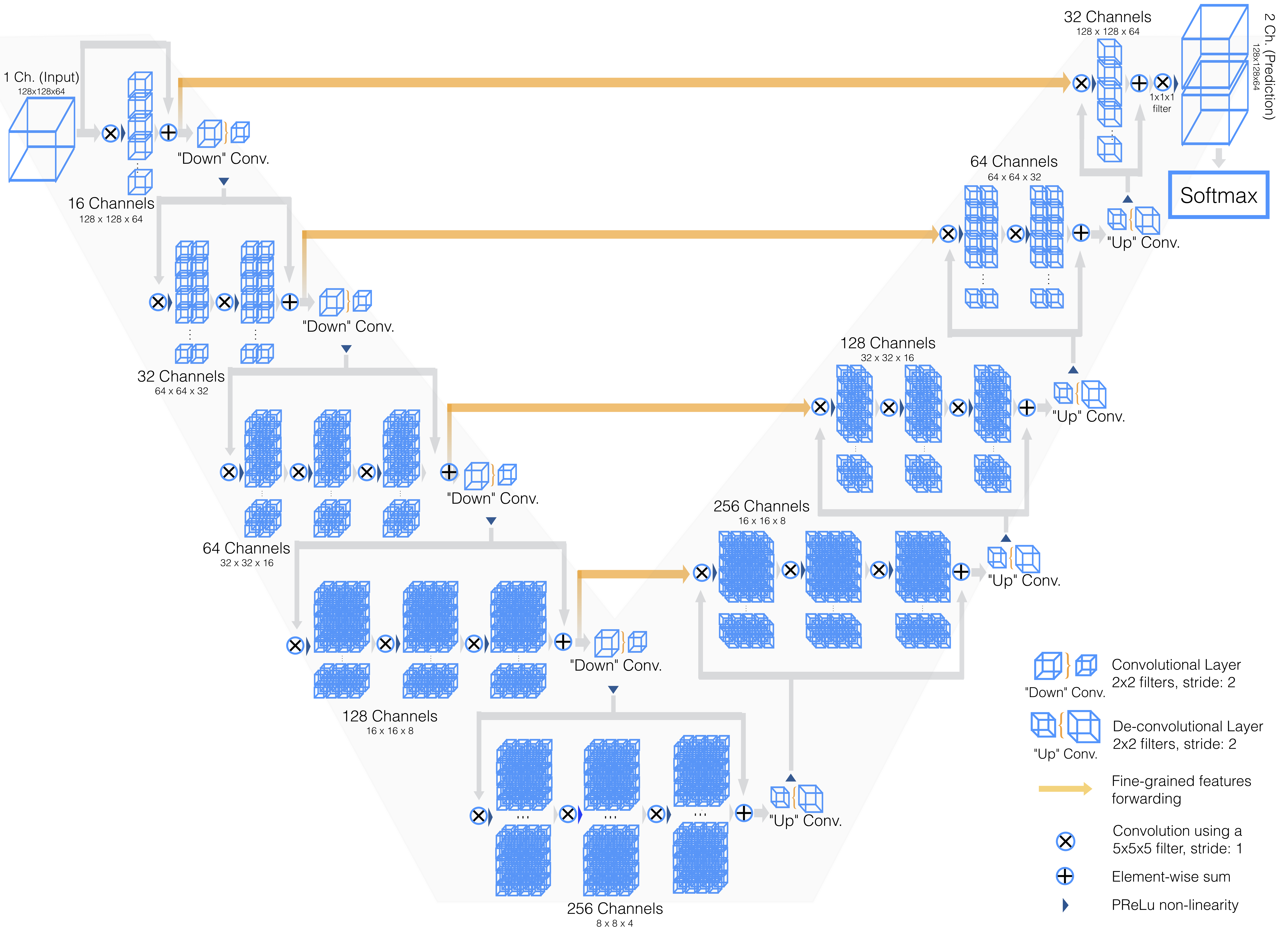} 	
\caption{V-Net: ``Schematic representation of our network architecture.'' ``...processes 3D data by performing volumetric convolutions.'' \cite{Milletari2016}} 
\label{fig:VnetImage} 
\end{figure}

Milletari et al. \cite{Milletari2016}  combined the above ideas in ``V-Net'', a 3D U-net with residual blocks applied to the task of 3D prostate MRI segmentation (see Figure \ref{fig:VnetImage}). 

The integration of greater spatial context and residual learning led to remarkable performance benefits, being on par with the then SoTA model on the ``PROMISE 2012'' challenge dataset at a reduced training convergence time common to residual networks. 
Unlike U-net \cite{Ronneberger2015} and 3D U-net \cite{Cicek2016}, the authors eschew batch normalization and follow the increasingly common trend of eliminating pooling layers, performing downsampling via convolutions kernels of size $2x2x2$ and a stride of two. They also performed segmentation on the entire image patch as opposed to previous works which only segmented the central section of the image patch. 

Another major contribution of the authors was the proposal of a \textit{soft DICE loss} which they used in their loss function in an attempt to directly optimize the network for segmentation accuracy. This version led to ~13\% greater performance than one trained using multinomial logistic loss with sample weighting. The resulting segmentation maps were not only more accurate, but also smoother and more visually pleasing.  

$LL_{dice}$ is a soft DICE loss applied to the  decoder output $p_{pred}$ to match the segmentation mask $p_{true}$:
   \begin{equation}
  \label{eq:dice}
  LL_{dice}  = \frac{2\times\sum_i^N p_{true} \times p_{pred} }{\sum_i^N p_{true}^2 + \sum_i^N p_{pred}^2 }   
  \end{equation}
where summation is voxel-wise.

\subsubsection{\textbf{V-net with Autoencoder Regularization}}
\begin{figure}[ht] 
    \centering
    \includegraphics[clip=true, trim=0pt 0pt 0pt 0pt, width=0.5\textwidth]{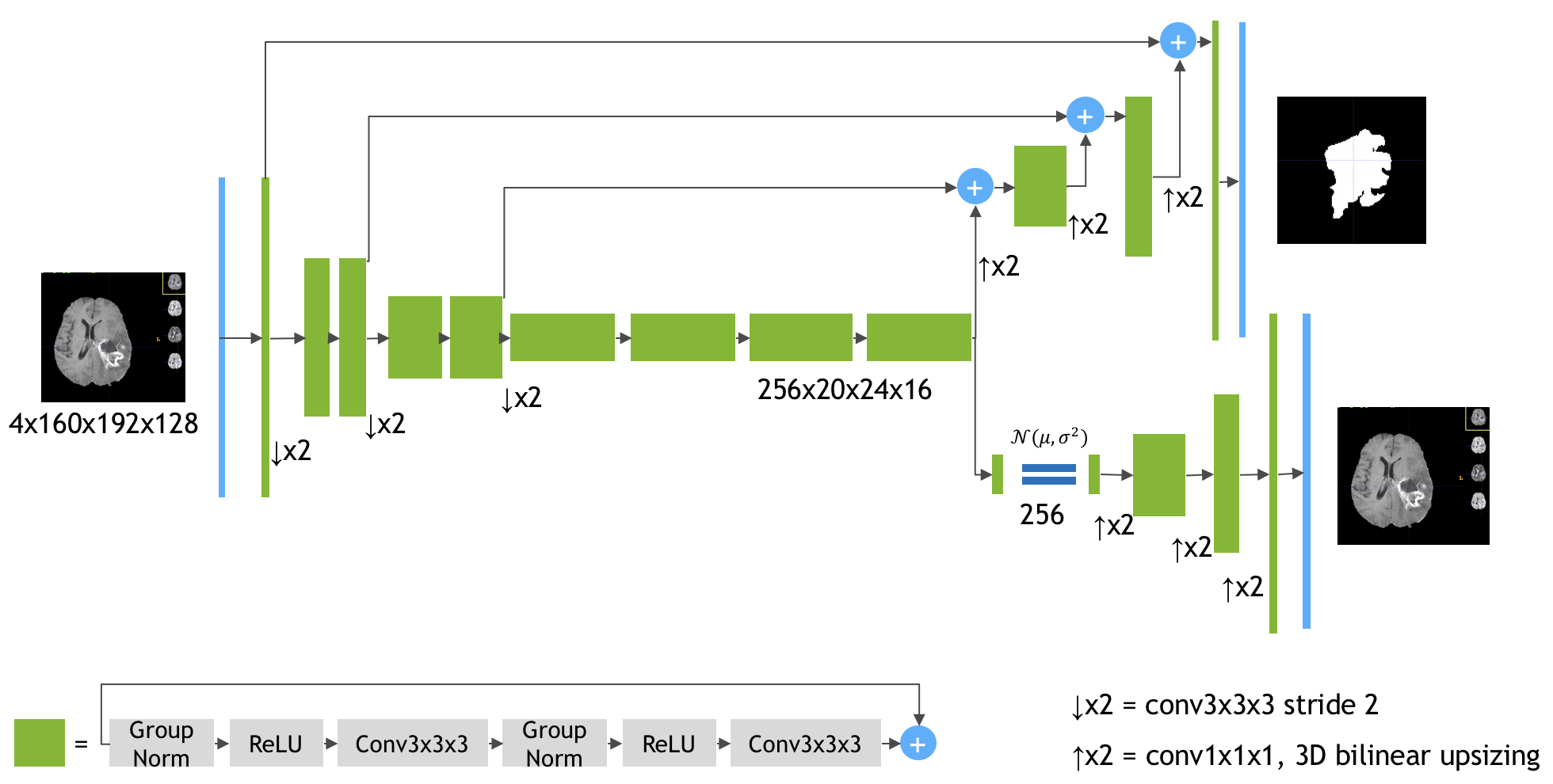}
    \caption{V-Net with autoencoder regularization: ``Schematic visualization of the network architecture.  Input is a four channel 3D MRI crop, followed by initial $3x3x3$ 3D convolution with 32 filters. Each green block is a ResNet-like block with the GroupNorm normalization. The output of the segmentation decoder has three channels (with the same spatial size as the input) followed by a sigmoid for segmentation maps of the three tumor subregions (WT, TC, ET).  The VAE branch reconstructs the input image into itself, and is used only during training to regularize the shared encoder.'' \cite{Myronenko2018}}
    \label{fig:myronenko_arch}
\end{figure}

Myronenko \cite{Myronenko2018} devised the current SoTA for 3D MRI brain tumor subregion segmentation and won the Medical Image Computing and Computer Assisted Intervention (MICCAI) Multimodal Brain Tumor Segmentation (BraTS) 2018 challenge. The author extended the V-net model by emphasizing the V-Net as an auto-encoder and imposing regularization contraints during training via a variational auto-encoder (VAE) \cite{Kingma2013} decoder branch attached to the encoder layer which bifurcated output to both the segmentation decoder branch and the VAE decoder branch (see Figure \ref{fig:myronenko_arch}). They then were able to leverage the KL divergence and L2 loss of the VAE branch in addition to a soft DICE loss (see Equation \ref{eq:dice}) \cite{Milletari2016} of the segmentation decoder branch in a composite loss function to both regularize the encoder and impose additional constraints. This had the effect of ensuring that features learned in the layers prior to the upsampling portion of the net minimized reconstruction error; in other words, biasing learned features to those that are the most salient and independent. The VAE branch was only used during training and was removed at test time. 

The output of the segmentation decoder branch is a direct segmentation map in the form of a three-channel image with spatial dimensions identical to the input image and each channel corresponding to one of three tumor subregion classes (i.e., enhancing tumor-core, whole tumor, tumor core).  

Another performance driver was their use of group normalization \cite{Wu2018}, being especially prudent given that the author forewent the use of multiple samples per batch (i.e., used a batch size of 1) in favor of maximizing the input image crop size, precluding the use of batch normalization. 

The aforementioned maximization of input crop size enabled the use of an extremely large input ($160x192x128$) relative to the original image size ($240x240x155$ for all samples). 
This is in contrast to the much smaller input used in EMMA \cite{Kamnitsas2018} ($64x64x64$), the prior year's SoTA approach, and No New-Net \cite{Isensee2018} ($128x128x128$), the current year's \nth{2} place method which incidentally was simply a 3D U-net \cite{Cicek2016} with the larger crop size (in addition to minor training differences and post-processing) being the only notable modification.  

These results seem to support the idea that the amount of spatial information has a much a greater impact on segmentation performance than complicated architectural modifications and pre-/post-processing techniques. Indeed, the author noted that experiments conducted with smaller input sizes, larger batch sizes, and the use of batch normalization resulted in worse performance. Experiments utilizing sophisticated data augmentation techniques (i.e., random histogram matching, affine image transforms, and random image filtering) showed no performance benefits. Tuning the network's segmentation results with conditional random fields showed mixed results, improving performance in some cases while reducing it in others. Performing test-time data augmentation by applying the neural network to 8 mirror flips of the 3D image and averaging realigned segmentation results led to an insignificant performance improvement. Ensembling 10 separately trained version of the model produced a roughly 1\% performance improvement.

The author also found that increasing the amount of information at each layer (i.e., number of learned filters) was able to consistently improve performance while increasing depth did not result in any performance gains, lending support to the idea that more salient low- to mid-level feature representations exist relative to higher-level feature spaces in the domain of medical imaging. This theory is further reinforced by \cite{Li2018} which was simply a U-Net \cite{Ronneberger2015} with eight additional convolutional layers immediately following the input layer that surpassed an ensemble of U-nets on the 2017 MICCAI WMH challenge.

\section{Deep Learning for Ultrasound: Applications, Implications, and Challenges}
\label{ultrasound}

Ultrasound is the most widely used modality in medical imaging but among the least researched in terms of automated analysis, possibly due to the fact that areas of clinical significance are fewer and less severe than, say, MRI, CT, or X-ray, which are generally reserved for the diagnosis of serious, often life-threatening pathologies. In contrast, ultrasound has historically being used chiefly in obstetrics, though it in theory could be used to support clinicians in a wide array of applications. Indeed, it is now often used to diagnose pathologies in parts of the body such as the heart, lung, prostate, thyroid, and breast; in image-assisted surgical procedures; and in point-of-care diagnostic pipelines in emergency medicine.   

Ultrasound is an imaging modality with particular clinical significance, being a safe, portable, relatively low-cost real-time diagnostic tool. It can be quickly and easily deployed in the field, making it especially valuable in disaster response scenarios and areas without adequate access to well-equipped medical facilities. Point-of-care ultrasound has been proven to provide faster, more precise diagnoses \cite{plummer1998emergency, Rodgerson2001}, reduce procedural complications \cite{Barnes2005}, and decrease time-to-treatment \cite{Bauman2009}. In combination with its use of non-ionizing radiation and fairly non-invasive application, it is afforded the unique ability to image ``any anatomy often, anywhere'', a feature that may otherwise be contraindicated, infeasible, or even impossible.  

Unfortunately, it possesses unique considerations absent in other modalities. Given the nature of ultrasonography and the typically free-hand nature of ultrasound image acquisition, image quality is highly dependent on the particular device, device settings, and acquisition technique used. This leads to the presence of speckle noise, artifacts, a greater emphasis on operator expertise, and significant variability between, and even within, observers. Additionally, 
ultrasound image interpretation relies on being able to dynamically investigate the same anatomical areas at different viewing angles in real-time, rather than over a set of static images obtained in advance.  These factors result in the need for a high level of expertise to properly acquire and interpret ultrasound images, a major barrier to full adoption by clinicians in all applicable areas.

Ultrasound technicians require much more training relative to other types of radiologists, training which necessarily must include exposure to a wide variety of pathologies across a wide variety of patients in a format that enables dynamic investigation, historically in-person workshops, with the confluence of these factors resulting in fewer opportunities to receive this training. Indeed, a 2018 study found that 84\% of the physician assistants surveyed believed that the training they received during their clinical rotations was insufficient preparation for clinical practice \cite{Barnett2018}. Among the many potential solutions to achieving sufficient ultrasonography expertise, increasing access to effective training opportunities via computer-based simulation platforms and developing automated image analysis systems to assist radiologists are two of the more promising areas. 

\subsection{Prospective Solutions}

\subsubsection{\textbf{Computer-based Simulation}}
Hardware and software platforms for ultrasound simulation have been proposed to directly empower radiologists-in-training and address the growing need for accessible ultrasound training \cite{Lewiss2014}. Of the many commercially available platforms, the SonoSim SonoSimulator in particular is among the most popular \cite{Truong2016, Rowley2019}. It has been shown to be as effective as live model training at teaching image acquisition \cite{Silva2016} and more effective for image interpretation \cite{Chung2013} with particular implications for urgent and logistically challenging scenarios, namely disaster-response training which can be greatly catalyzed by the focused assessment with sonography in trauma (FAST) protocol \cite{Paddock2015}. The major advantage of the SonoSim SonoSimulator over other platforms is its high-fidelity simulation, usage of real patient data (as opposed to most other commercially available platforms which use synthetically generated data due to the difficulties in obtaining high-quality in-vivo data), and thousands of cases with a wide array of pathologies. This not only allows for repeated, realistic training on typical cases but also on rare but serious conditions which, due to their infrequence, are likely to be underdetected when presented in the clinical setting \cite{Lewiss2014}.

\subsubsection{\textbf{Automated Image Analysis Algorithms}}
In conjunction with greater access to high-quality training, automated image analysis could increase clinical effectiveness and lower the amount of time, expertise, and cognitive resources required of sonographers by facilitating image interpretation through capabilities such as anatomy detection, classification, and semantic segmentation; object disambiguation with noise and artifact reduction through ``salient signal recovery'' \cite{Zosa2018}; and directional guidance/feedback in image-assisted interventions.

\subsubsection{\textbf{A Dual Pathway Approach}}
Finally, the combination of automated image analysis into an ultrasound training platform could catalyze expertise acquisition while reducing inter- and intra-observer variability by providing a high-quality, standardized training experience. This could be achieved via the integration of automated analyses that would be used in the clinic or by scaffolding learning with tools tailor-made to complement a didactic pedagogy.

\subsection{Segmentation Challenges in Ultrasound}
As segmentation is a necessary component of many medical image analysis workflows, as well as an end in itself, research in this area is the logical first step in achieving robust automated image analysis. Unfortunately, there is still a lack of research in ultrasound image segmentation relative to, say, MRI or CT. 

This is compounded by (and most likely due to) the fact that sonography produces images that are harder to analyze given inherent inhomogeneities due to speckle intensity, a low signal-to-noise ratio, shadows, and edge dropout. In addition, there is wide variance in image and target features across devices, anatomies, and patients (i.e., significant appearance and shape variations) with a priori anatomical knowledge influencing the images acquired (i.e., acquisition plane and location) and greatly informing the segmentation task (e.g. determining whether an image region is a shadow, artifact, or anatomy). 

This implies that, in contrast to other high-fidelity imaging modalities where strong segmentation results have been achieved with the incorporation of a relatively minor amount of a priori information or spatial contextual information, the idiosyncratic challenges of ultrasound image analysis necessitate the inclusion of a higher degree of both sources of information to produce expert-level segmentation algorithms.

\subsection{Recent Deep Learning-based Approaches}
Given that neural network performance is dictated by the quality of the high-level abstractions gleaned from raw image features, the low-fidelity image features produced by ultrasonography have generally hampered the development of strong deep learning models. In contrast, classic high-level methods take more information into account and are able to model the geometry and physics of the target anatomy. Unfortunately, these models have historically been laborious to implement as modeling of those parameters necessitates feature engineering which in turn relies on a priori knowledge about the speckle patterns of regions of interest as well as organ geometry \cite{Nandamuri2019}. In response, researchers have investigated using deep neural networks, especially in 3D space, to automatically extract features to drive classical methods in a hybrid framework.

\subsubsection{\textbf{Sparse Adapative Neural Networks (SADNN)}}
\begin{figure}[ht] 
    \centering
    \includegraphics[clip=true, trim=0pt 0pt 0pt 0pt, width=0.45\textwidth]{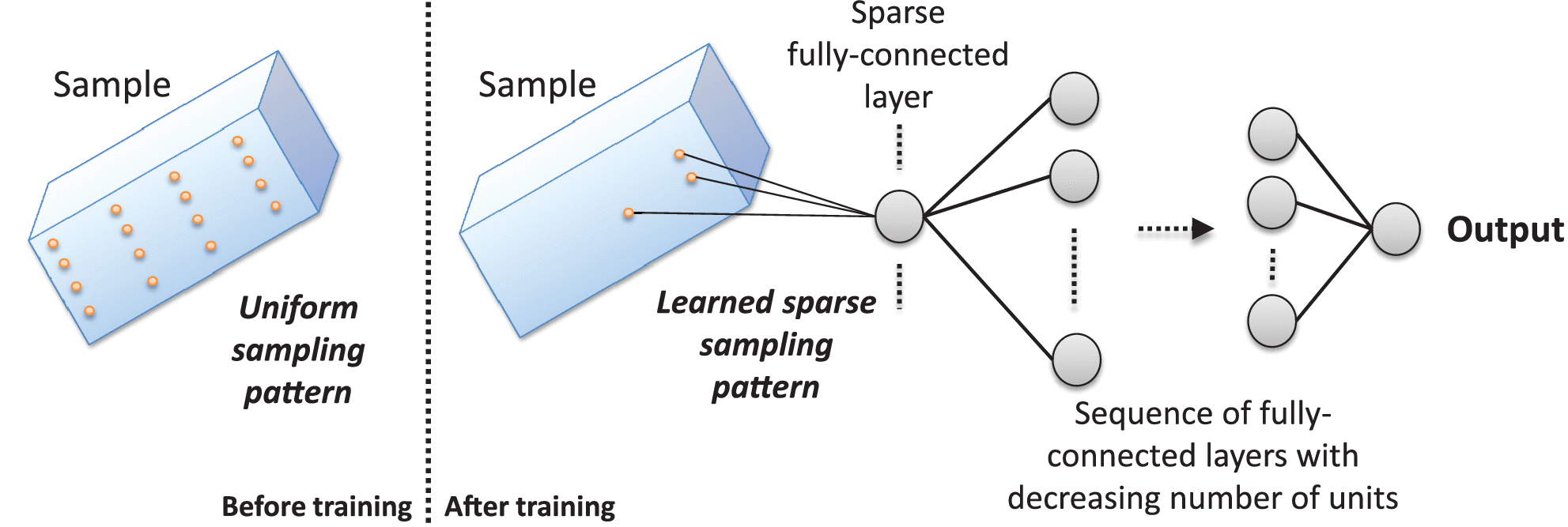}
    \caption{SADNN: ``Visualization of the difference between uniform/handcrafted feature patterns and self-learned, sparse, adaptive patterns'' \cite{Ghesu2016}}
    \label{fig:marginal_sampling}
\end{figure}

\begin{figure}[ht] 
    \centering
    \includegraphics[clip=true, trim=0pt 0pt 0pt 0pt, width=0.45\textwidth]{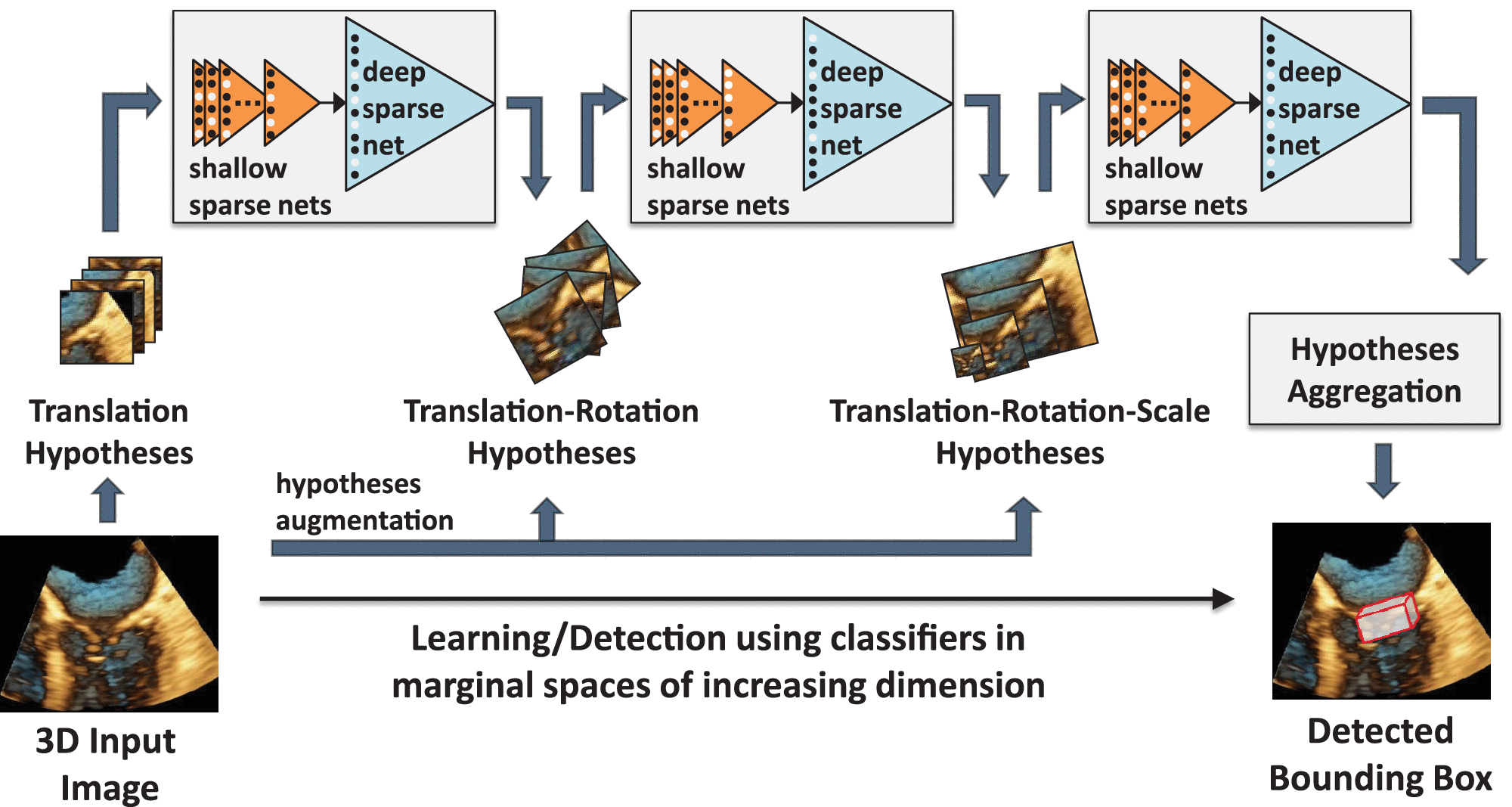}
    \caption{SADNN: ``Schematic visualization of the marginal space deep learning framework. The black/white dots encode the sparse sampling patterns.'' \cite{Ghesu2016}}
    \label{fig:marginal_boxes_hyp}
\end{figure}

\begin{figure}[ht] 
    \centering
    \includegraphics[clip=true, trim=0pt 0pt 0pt 0pt, width=0.45\textwidth]{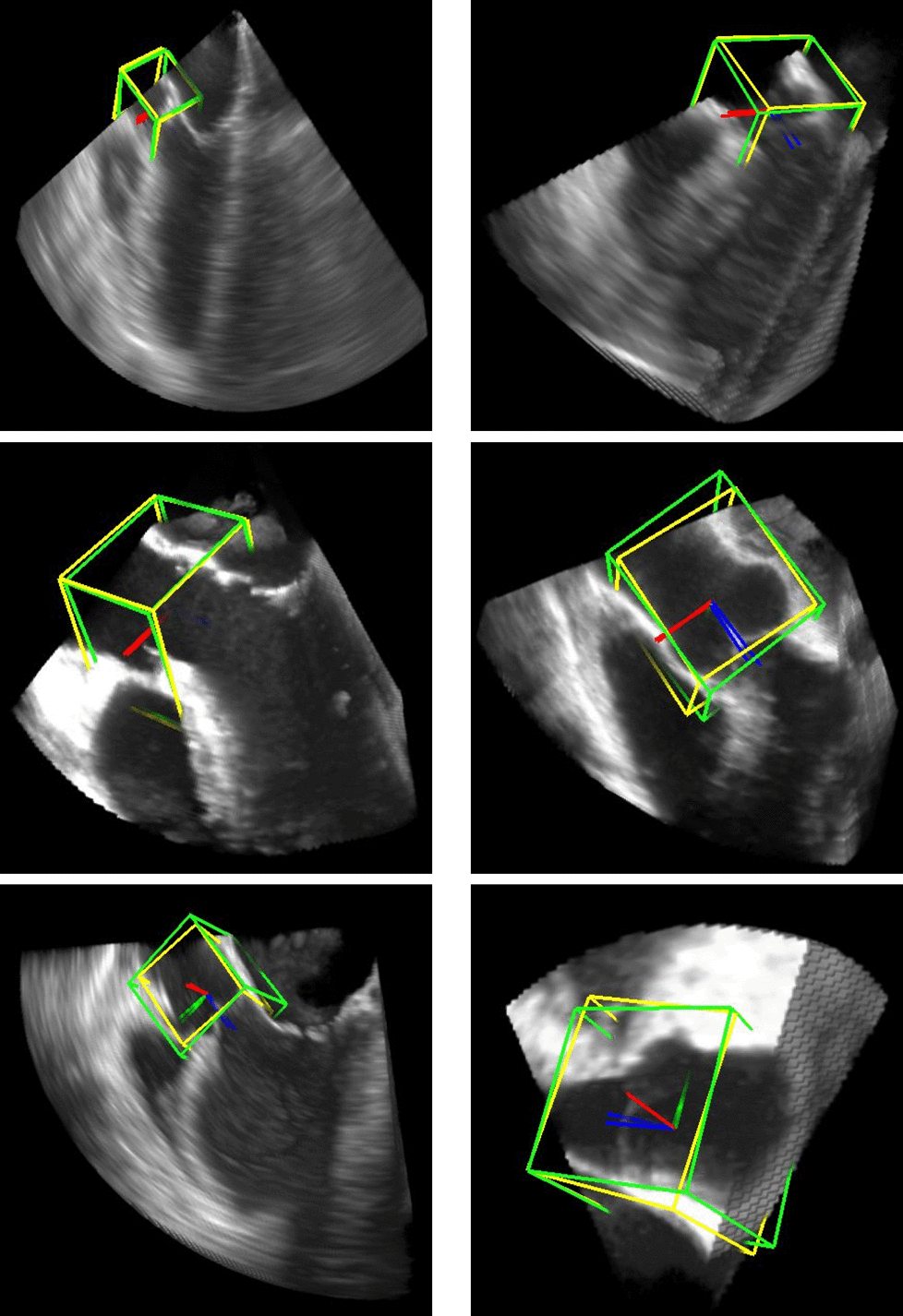}
    \caption{SADNN: ``Example images showing the detection results for different patients from the test set. The detected bounding box is visualized in green and the ground truth box in yellow. The segments with the origin in the center of each box define the corresponding coordinate system. Note that as specified in the text, the 3D pose of the aortic valve (position, orientation and scale) is fully determined by the anatomy.'' \cite{Ghesu2016}}
    \label{fig:marginal_boxes_gen}
\end{figure}

\begin{figure}[ht] 
    \centering
    \includegraphics[clip=true, trim=0pt 0pt 0pt 0pt, width=0.45\textwidth]{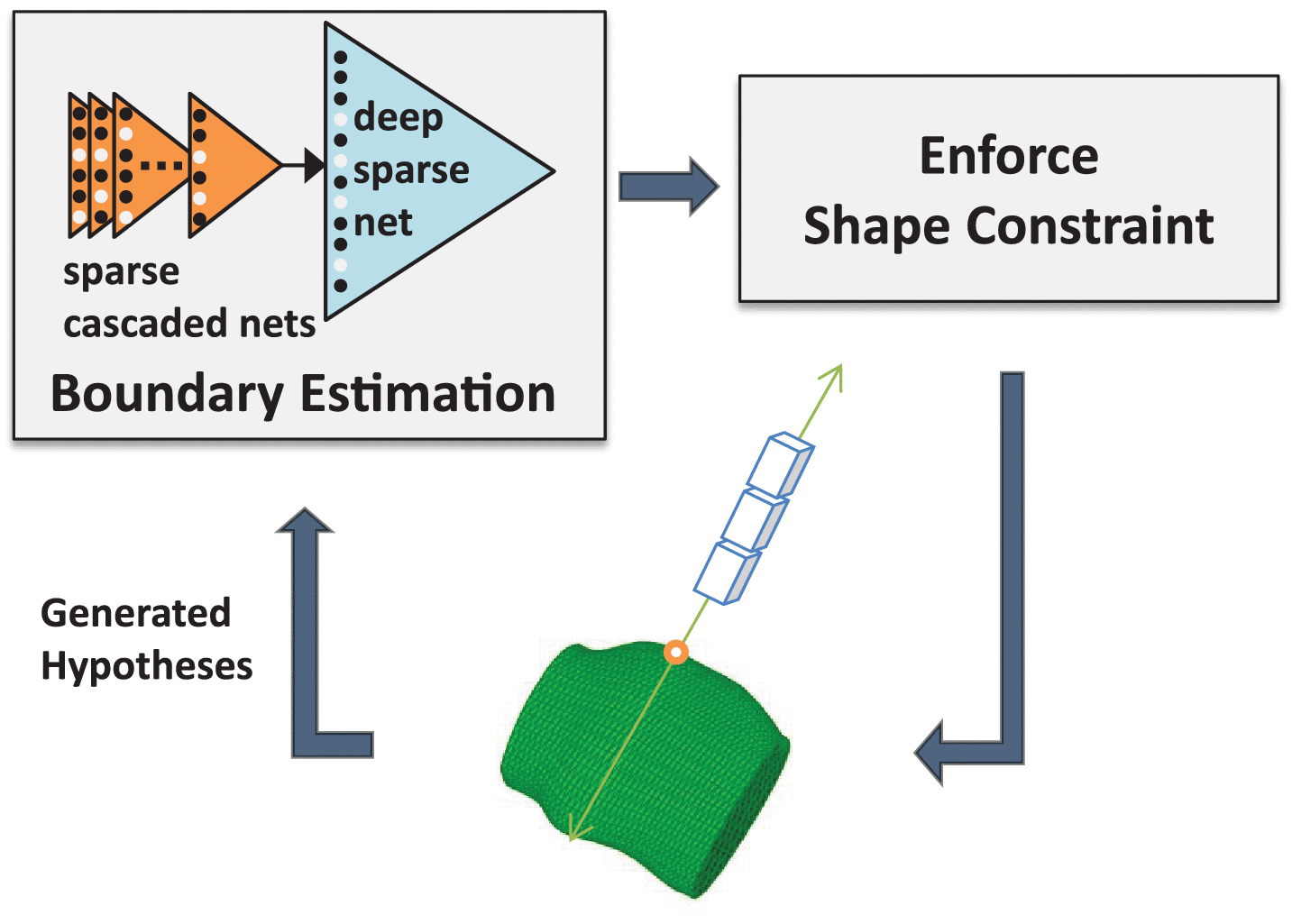}
    \caption{SADNN: ``Schematic visualization of the boundary deformation with SADNN. Starting from the current shape, the SADNN is aligned and applied along the normal for each point of the mesh, the boundary is deformed and projected under the current shape space. The process is iteratively repeated. The black/white dots encode the sparse patterns, learned in the cascaded shallow networks and deep boundary classifier.'' \cite{Ghesu2016}}
    \label{fig:marginal_active_shape}
\end{figure}

\begin{figure}[ht] 
    \centering
    \includegraphics[clip=true, trim=0pt 0pt 0pt 0pt, width=0.45\textwidth]{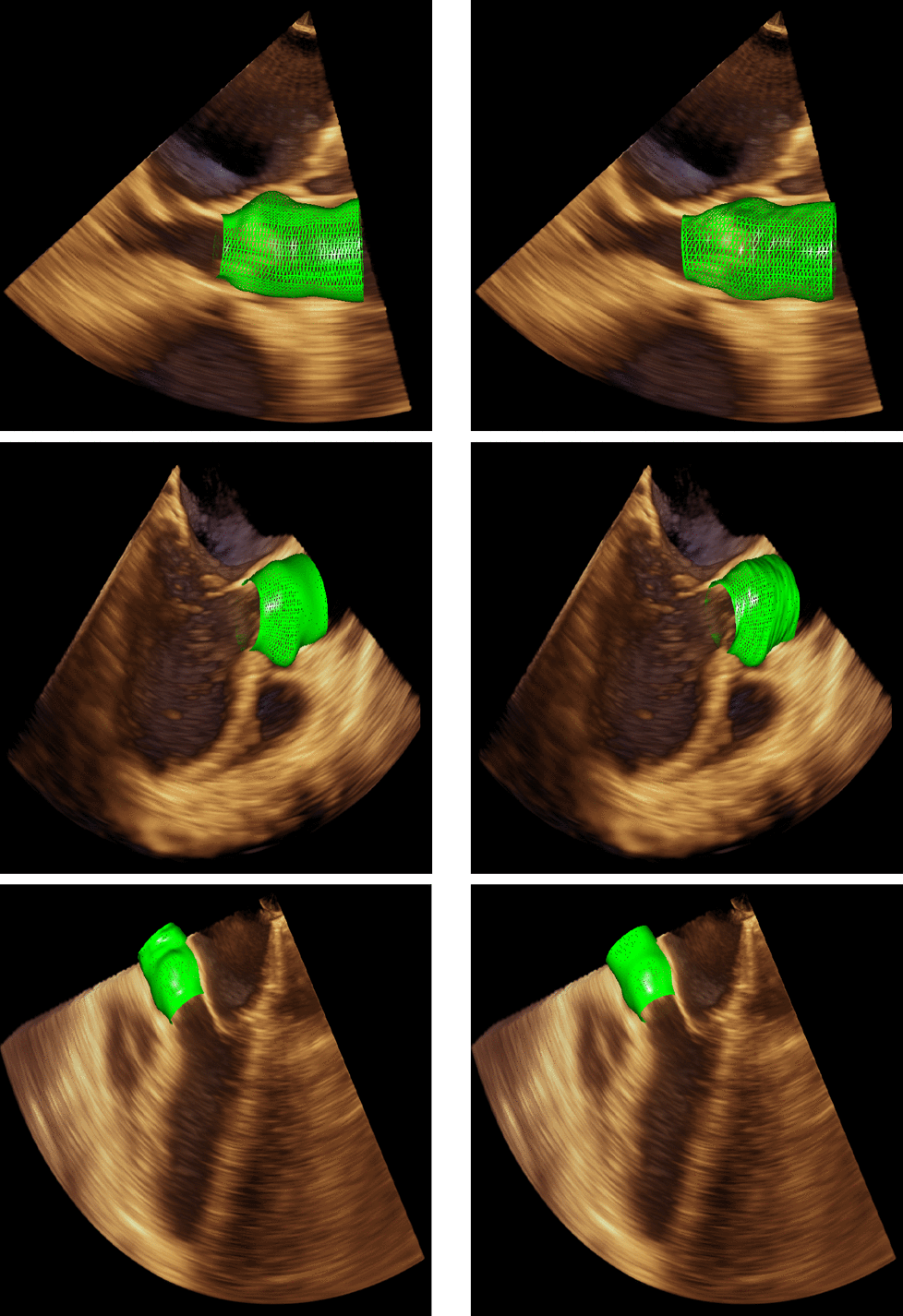}
    \caption{SADNN: ``Example images showing the aortic valve segmentation results for different patients from the test set, using our proposed method. Each row of images corresponds to a different patient, the left image represents the ground truth mesh, while the right image shows the detected mesh.'' \cite{Ghesu2016}}
    \label{fig:marginal_final}
\end{figure}

Ghesu et al. \cite{Ghesu2016} proposed a sparse adaptive neural network (SADNN). SADNN used marginal space learning to enable a fully-connected network to efficiently utilize spatial information and drive an active shape model for aortic valve segmentation 
(see Figures \ref{fig:marginal_sampling} - \ref{fig:marginal_final} for a visual overview), 
achieving over 45.2\% mean position error improvement over the previous marginal space learning state of the art method, and at the same run-time.

\subsubsection{\textbf{fCNN-enhanced 3D Snake}}
\begin{figure}[ht] 
    \centering
    \includegraphics[clip=true, trim=0pt 0pt 0pt 0pt, width=0.5\textwidth]{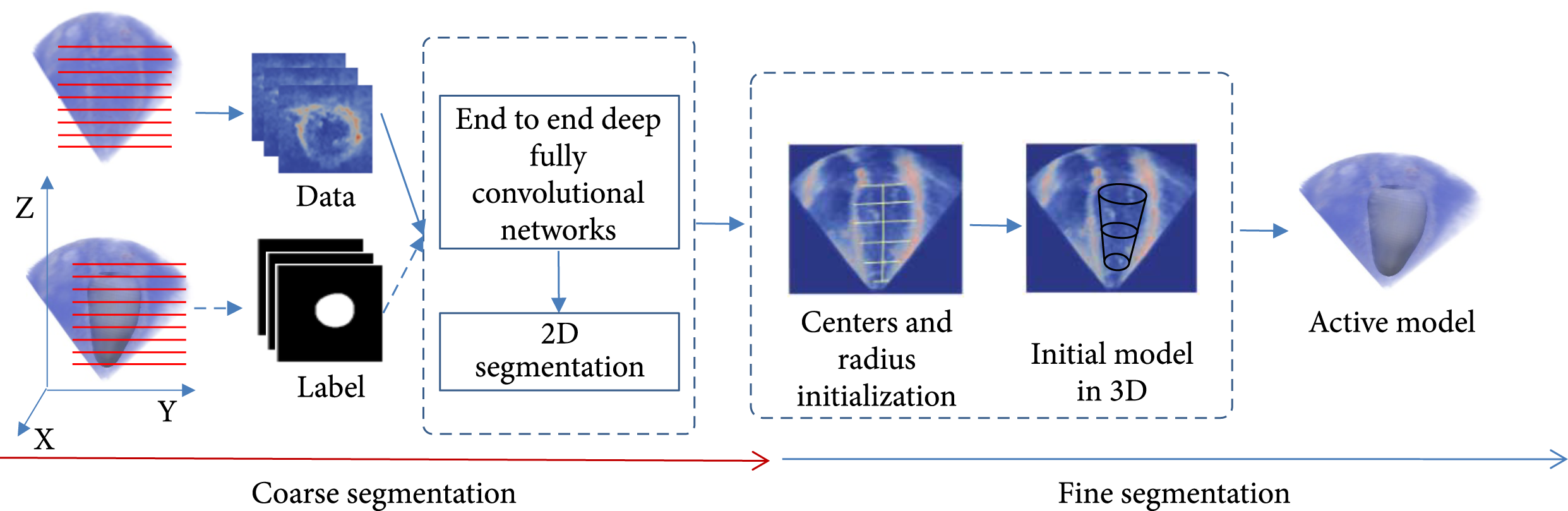}
    \caption{fCNN-enhanced 3D Snake \cite{Dong2018}}.
    \label{fig:c2f_snake}
\end{figure}

Similarly, Dong et al. \cite{Dong2018} utilized a fCNN with a deformable model (see Figure \ref{fig:c2f_snake}) to segment the adult LV. Their paradigm employed a 2D fCNN on 3D image slices, using the fCNN to generate a coarse segmentation and initialize the deformable model, a 3D snake, which subsequently was used to produce a fine segmentation. This approach again outperformed either method alone and also surpassed the SoTA U-Net on the Challenge on Endocardial Three-dimensional Ultrasound Segmentation (CETUS) dataset \cite{Bernard2016}.

\subsubsection{\textbf{fCNN-enhanced 3D Atlas}}
\begin{figure}[ht] 
    \centering
    \includegraphics[clip=true, trim=0pt 0pt 0pt 0pt, width=0.5\textwidth]{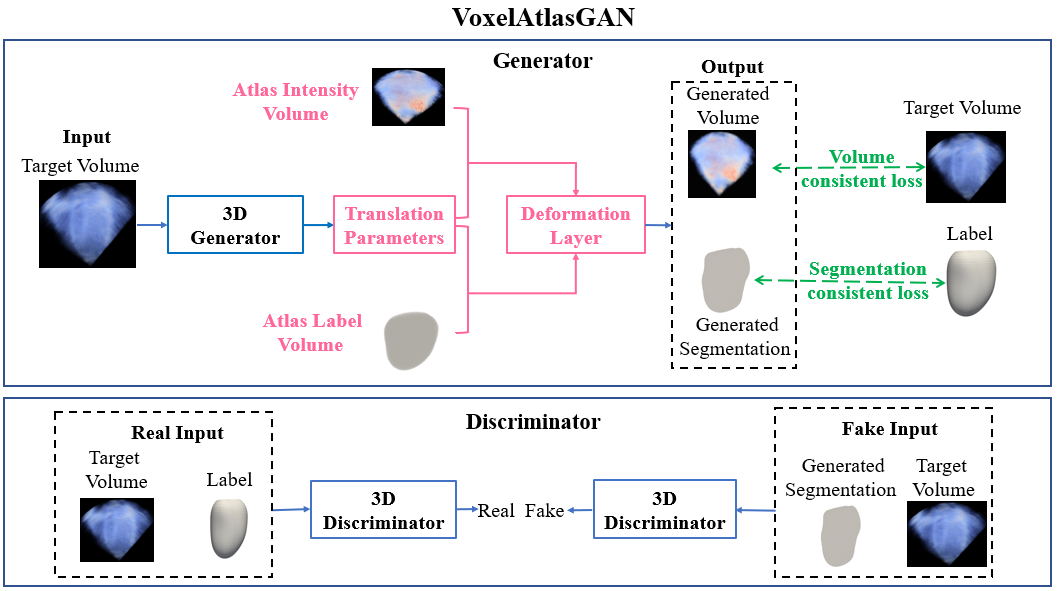}
    \caption{fCNN-enhanced 3D Atlas \cite{Dong2018}}.
    \label{fig:c2f_VoxelAtlasGAN}
\end{figure}

Dong et al. \cite{Dong2018a} also integrated an atlas-based method with a 3D fCNN in a conditional generative adversarial paradigm to learn the transformation parameters of the atlas (see Figure \ref{fig:c2f_VoxelAtlasGAN}), achieving a 95.3\% DICE score, 6\% higher than the SoTA V-Net on the CETUS dataset \cite{Bernard2016}.

\subsubsection{\textbf{SumNet}}
\begin{figure}[ht] 
    \centering
    \includegraphics[clip=true, trim=0pt 0pt 0pt 0pt, width=0.5\textwidth]{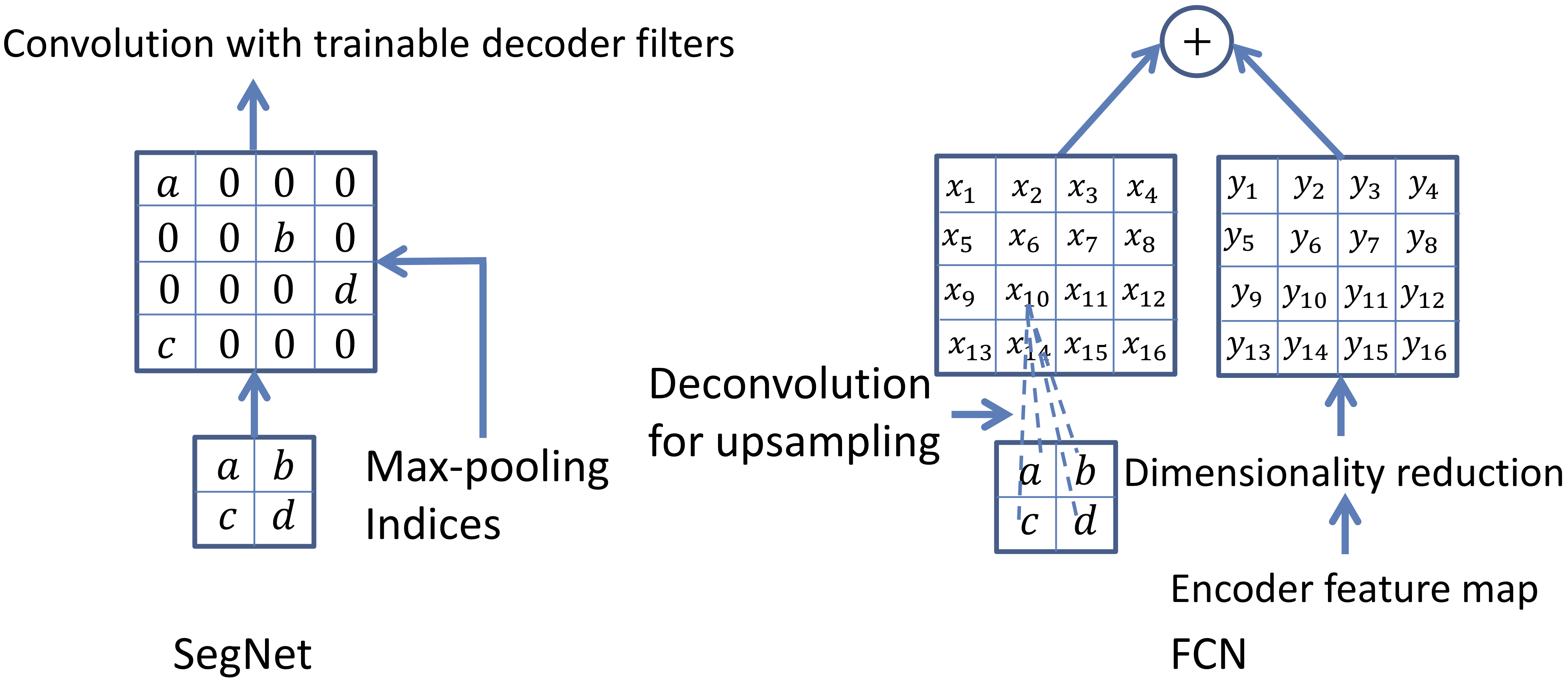}
    \caption{SumNet:\protect\footnotemark[2] ``An illustration of SegNet and FCN \cite{Long2015} decoders. $a,b,c,d$ correspond to values in a feature map. SegNet uses the max pooling indices to upsample (without learning) the feature map(s) and convolves with a trainable decoder filter bank. FCN upsamples by learning to deconvolve the input feature map and adds the corresponding encoder feature map to produce the decoder output. This feature map is the output of the max-pooling layer (includes sub-sampling) in the corresponding encoder. Note that there are no trainable decoder filters in FCN.'' \cite{Badrinarayanan2017}}
    \label{fig:SegNet_PoolingIndexes}
\end{figure}

While these results are provocative, a recent notable exception to the aforementioned trend of hybrid methods and 3D information is SumNet \cite{Nandamuri2019}, a SegNet-based \cite{Badrinarayanan2017} neural network, which uses max pooling indexes to improve information transfer during upsampling (see Figure \ref{fig:SegNet_PoolingIndexes}), to tackle the task of intravascular ultrasound segmentation (IVUS) and thyroid gland segmentation. The network takes in 2D slices from a single plane, computes the segmentation mask over each slice, and concatenates these slices back together to form the complete 3D volumetric segmentation. They improved on the SoTA in both tasks, achieving a 93\% and 92\% DICE score on the 2011 MICCAI IVUS dataset and a publicly available thyroid dataset \cite{Wunderling2017}, respectively.

\subsection{Implications of Interpretability}
Hybrid approaches have been shown to offer many advantages. Not only have they set record performance, they also address the criticism of lack of algorithm interpretability that plague neural network-only solutions by providing interpretability at the layer of the high-level method. This is especially important in medical applications as understanding what the model is doing can provide better guarantees on performance and help identify scenarios to which the model is best suited. As physicians rely on the outputs of these models to make critical clinical decisions, it is of paramount importance that they are deployed only in the appropriate clinical contexts and that clinicians understand the model to the degree that helps them best decide how to integrate it into their practice.
\subsection{Summary}
The most recent wave of cutting edge research demonstrates that the integration of neural networks that derive salient voxel-derived features with high-level methods to leverage a priori information is an effective approach. However, extending this paradigm to incorporate the full spatial extent of the anatomy in question via some form of 3D neural network is still a nascent area of research.  In addition, recent findings suggest that we have yet to discover the most effective neural network-only solution. Consequently, more research is needed on both more effective neural networks and hybrid schemes.

Given the promising initial results, difficulties inherent in ultrasound image analysis, and consequent high potential impact of expert-level computer-assisted workflows, this branch of research will soon define the gold standard of modern ultrasound image analysis methods, and potentially the entire field of medical image analysis. In combination with digital training platforms, these methods will reduce the historical barriers to ultrasound usage by catalyzing ultrasound training or directly assisting clinicians in their practice, enabling manifold novel uses across a variety of clinical practices, greater success of image-assisted procedures, and greater access to point-of-care diagnostics which are are crucial for disaster and trauma scenarios.

\section{Future Directions}
\label{future}

In addition to the architectures and approaches discussed, there are also multiple avenues for improvement via integration of other ideas from related tasks, such as:
\begin{enumerate}
    \item Pairing raw input images with hand engineered features based on a priori information as in \cite{Dollar}.
    \item Introducing slightly more sophisticated, extensible modules to complement notions of network depth and width while still keeping overall network design simple and minimizing hyperparameter complexity. Methods utilizing this approach have already shown SoTA performance on related tasks such as image classification \cite{Xie2017} and image enhancement \cite{Zhang2018}.
    \item Utilizing hybrid architectures which only apply 3D convolutions over the feature space most aided by their inclusion in order to alleviate the exponential increase in computation and memory in addition to the overfitting predilection of pure 3D networks. This may be application-specific, with 3D convolutions sometimes being more useful when applied to high-level features \cite{Xie2018} and sometimes when applied to low-level features \cite{Shan2018}.  This can be further aided by the decomposition of 3D convolutions on spatiotemporal data to separate 2D convolutions over the spatial and temporal axes, which has been shown to yield higher performance in addition to parameter and computation reduction \cite{Xie2018}.
\end{enumerate}

\footnotetext[2]{Figure from \cite{Badrinarayanan2017}}

\section{Conclusion}
\label{conclusion}
Medical image segmentation algorithms have seen many improvements in the past few decades. From the proposal of low-level edge-, region-, and classification-based methods, to high-level atlas-based methods and deformable parametric and geometric models, to the recent breakthroughs utilizing deep neural networks, automated algorithms have steadily approached expert-level performance. 

Many recent architectures applied to biomedical segmentation tasks achieve success due to their ability to simultaneously integrate image features at different scales, building salient abstract feature representations that lead to the production of high-quality segmentation masks. They have made use of simple architectural improvements such as the use of residual learning and the integration of 3D context, as well as major insights such as the regularization of layers through hybrid architectures and loss functions that directly minimize the segmentation map error metric.  These architectures have defined the current gold standard on those tasks. 

While high-quality image analysis algorithms will most certainly be useful across all medical imaging modalites, some may experience a greater benefit than others. For example, ultrasonography is extremely useful in a wide variety of scenarios but has not been able to achieve full adoption across relevant clinical specialties due to challenges idiosyncratic to ultrasound image acquisition and interpretation. Automated expert-level ultrasound image analysis systems have the potential to overcome these challenges, facilitating clinical expertise acquisition while at the the same time making a much wider impact due to the accessibility and manifold potential applications of medical ultrasound. There is a growing body of research in this area, though it still lags behind research in other modalities such as CT and MRI. 

Finally, across all medical imaging modalities, there are various possible optimizations and improvements that have been developed for related tasks that have the potential to improve neural network-based segmentation performance. 

While neural networks have significantly narrowed the performance gap between human experts and automated algorithms, the gap still exists. In addition, at present there is no single general framework which applies to segmentation tasks across all anatomies or imaging modalities. Instead, approaches must still be tailored to the specific domain, task, and, sometimes even, dataset. Consequently the task of automated medical image segmentation is far from solved and much work remains to be done to enable the development of automatic expert-level segmentation algorithms.




\section*{Conflict of interest}
The author is currently employed by SonoSim, Inc. in their Research \& Development Division.

\section*{Acknowledgments}

The author wishes to thank his Doctoral Research Mastery Exam Committee Chair Dr. Lawrence Saul as well as fellow committee members Drs. Hao Su and Garrison Cottrell.
In particular, he thanks Dr. Saul for his support in preparing for the research examination, Dr. Su for his candor and career advice early in the author's academic career, and Dr. Cottrell for his positivity and patience.

He also wishes to thank Dr. Michael Hazoglou and Monique Narboneta for critiquing countless iterations of the paper and oral presentation; SonoSim colleagues Drs. Matthew Wang, Kresimir Petrinec, and Gabriele Nataneli, CTO, for their counsel, mentorship, and constant moral support throughout the author's research career; and Dr. Eric Savitsky, M.D., CEO, for his belief in the author's potential as a researcher and consequent support and encouragement to explore the application of the author's research at SonoSim, Inc. 

It is only by God's grace made manifest through the help of the aforementioned and various unmentioned individuals that the author can claim any semblance of success, and thus reserves no credit for himself. This work represents a cumulative effort and the author wishes to once again extend his heartfelt thanks and acknowledgement to every person who had a part in it. 



\bibliographystyle{IEEEtran}
\bibliography{references}

\begin{thebibliography}{10}
\providecommand{\url}[1]{#1}
\csname url@samestyle\endcsname
\providecommand{\newblock}{\relax}
\providecommand{\bibinfo}[2]{#2}
\providecommand{\BIBentrySTDinterwordspacing}{\spaceskip=0pt\relax}
\providecommand{\BIBentryALTinterwordstretchfactor}{4}
\providecommand{\BIBentryALTinterwordspacing}{\spaceskip=\fontdimen2\font plus
\BIBentryALTinterwordstretchfactor\fontdimen3\font minus
  \fontdimen4\font\relax}
\providecommand{\BIBforeignlanguage}[2]{{%
\expandafter\ifx\csname l@#1\endcsname\relax
\typeout{** WARNING: IEEEtran.bst: No hyphenation pattern has been}%
\typeout{** loaded for the language `#1'. Using the pattern for}%
\typeout{** the default language instead.}%
\else
\language=\csname l@#1\endcsname
\fi
#2}}
\providecommand{\BIBdecl}{\relax}
\BIBdecl

\bibitem{Liu2019}
\BIBentryALTinterwordspacing
S.~Liu, Y.~Wang, X.~Yang, B.~Lei, L.~Liu, S.~X. Li, D.~Ni, and T.~Wang, ``{Deep
  Learning in Medical Ultrasound Analysis: A Review},'' \emph{Engineering},
  vol.~5, no.~2, pp. 261--275, 2019. [Online]. Available:
  \url{https://doi.org/10.1016/j.eng.2018.11.020}
\BIBentrySTDinterwordspacing

\bibitem{Yang2016}
\BIBentryALTinterwordspacing
X.~Yang, L.~Yu, L.~Wu, Y.~Wang, D.~Ni, J.~Qin, and P.-A. Heng, ``{Fine-grained
  Recurrent Neural Networks for Automatic Prostate Segmentation in Ultrasound
  Images},'' in \emph{Proceedings of the Thirty-First AAAI Conference on
  Artificial Intelligence (AAAI-17) Fine-Grained}, 2016, pp. 1633--1639.
  [Online]. Available: \url{http://arxiv.org/abs/1612.01655}
\BIBentrySTDinterwordspacing

\bibitem{Nascimento2016}
\BIBentryALTinterwordspacing
J.~C. Nascimento and G.~Carneiro, ``{Multi-atlas segmentation using manifold
  learning with deep belief networks},'' in \emph{2016 IEEE 13th International
  Symposium on Biomedical Imaging (ISBI)}.\hskip 1em plus 0.5em minus
  0.4em\relax IEEE, apr 2016, pp. 867--871. [Online]. Available:
  \url{https://cs.adelaide.edu.au/{~}carneiro/publications/Isbi-2016.pdf
  http://ieeexplore.ieee.org/document/7493403/}
\BIBentrySTDinterwordspacing

\bibitem{Li2017}
Y.~Li, R.~Xu, J.~Ohya, and H.~Iwata, ``{Automatic fetal body and amniotic fluid
  segmentation from fetal ultrasound images by encoder-decoder network with
  inner layers},'' \emph{Proceedings of the Annual International Conference of
  the IEEE Engineering in Medicine and Biology Society, EMBS}, pp. 1485--1488,
  2017.

\bibitem{Ma2017}
J.~Ma, F.~Wu, T.~Jiang, Q.~Zhao, and D.~Kong, ``{Ultrasound image-based thyroid
  nodule automatic segmentation using convolutional neural networks},''
  \emph{International Journal of Computer Assisted Radiology and Surgery},
  vol.~12, no.~11, pp. 1895--1910, 2017.

\bibitem{Hafiane2017}
\BIBentryALTinterwordspacing
A.~Hafiane, P.~Vieyres, and A.~Delbos, ``{Deep learning with spatiotemporal
  consistency for nerve segmentation in ultrasound images},'' \emph{CoRR}, vol.
  abs/1706.0, 2017. [Online]. Available: \url{http://arxiv.org/abs/1706.05870}
\BIBentrySTDinterwordspacing

\bibitem{Zhang2017}
Y.~Zhang, M.~T. Ying, L.~Yang, A.~T. Ahuja, and D.~Z. Chen, ``{Coarse-to-Fine
  Stacked Fully Convolutional Nets for lymph node segmentation in ultrasound
  images},'' \emph{Proceedings - 2016 IEEE International Conference on
  Bioinformatics and Biomedicine, BIBM 2016}, pp. 443--448, 2017.

\bibitem{Singhal2017}
N.~Singhal, S.~Mukherjee, and C.~Perrey, ``{Automated assessment of endometrium
  from transvaginal ultrasound using Deep Learned Snake},'' \emph{Proceedings -
  International Symposium on Biomedical Imaging}, pp. 283--286, 2017.

\bibitem{Milletari2017}
\BIBentryALTinterwordspacing
F.~Milletari, S.-a. Ahmadi, C.~Kroll, A.~Plate, V.~Rozanski, J.~Maiostre,
  J.~Levin, O.~Dietrich, B.~Ertl-wagner, K.~B{\"{o}}tzel, and N.~Navab,
  ``{Hough-CNN : Deep learning for segmentation of deep brain regions in MRI
  and ultrasound},'' \emph{Computer Vision and Image Understanding}, vol. 164,
  pp. 92--102, 2017. [Online]. Available:
  \url{https://doi.org/10.1016/j.cviu.2017.04.002}
\BIBentrySTDinterwordspacing

\bibitem{Aslan2018}
\BIBentryALTinterwordspacing
E.~Aslan, N.~Dumlu, and Y.~S. Akgul, ``{Tongue contour extraction from
  ultrasound images using image parts},'' in \emph{2018 26th Signal Processing
  and Communications Applications Conference (SIU)}.\hskip 1em plus 0.5em minus
  0.4em\relax IEEE, may 2018, pp. 1--4. [Online]. Available:
  \url{https://arxiv.org/pdf/1605.05912.pdf
  https://ieeexplore.ieee.org/document/8404554/}
\BIBentrySTDinterwordspacing

\bibitem{Royan2012}
S.~Royan, \emph{{Oxford Handbook of Medical Imaging}}, 2012, vol.~85, no. 1018.

\bibitem{JainRameshandKasturiRangacharandSchunck1995}
B.~G. {Jain, Ramesh and Kasturi, Rangachar and Schunck}, \emph{{Machine
  Vision}}.\hskip 1em plus 0.5em minus 0.4em\relax New York, NY, USA:
  McGraw-Hill, Inc., 1995.

\bibitem{Norouzi2014}
\BIBentryALTinterwordspacing
A.~Norouzi, M.~S.~M. Rahim, A.~Altameem, T.~Saba, A.~E. Rad, A.~Rehman, and
  M.~Uddin, ``{Medical image segmentation methods, algorithms, and
  applications},'' \emph{IETE Technical Review (Institution of Electronics and
  Telecommunication Engineers, India)}, vol.~31, no.~3, pp. 199--213, 2014.
  [Online]. Available: \url{http://dx.doi.org/10.1080/02564602.2014.906861}
\BIBentrySTDinterwordspacing

\bibitem{Adams1994}
R.~Adams and L.~Bischof, ``{Seeded Region Growing},'' \emph{IEEE Transactions
  on Pattern Analysis and Machine Intelligence}, vol.~16, no.~6, pp. 641--647,
  1994.

\bibitem{Kass1988}
\BIBentryALTinterwordspacing
M.~Kass, A.~Witkin, and D.~Terzopoulos, ``{Snakes: Active contour models},''
  \emph{International Journal of Computer Vision}, vol.~1, no.~4, pp. 321--331,
  jan 1988. [Online]. Available:
  \url{http://link.springer.com/10.1007/BF00133570}
\BIBentrySTDinterwordspacing

\bibitem{Osher1988}
\BIBentryALTinterwordspacing
S.~Osher and J.~A. Sethian, ``{Fronts Propagating with Curvature Dependent
  Speed},'' \emph{Journal of computational physics}, vol.~79, no.~1, pp.
  12--49, 1988. [Online]. Available:
  \url{https://ntrs.nasa.gov/archive/nasa/casi.ntrs.nasa.gov/19880001113.pdf}
\BIBentrySTDinterwordspacing

\bibitem{Bengio2013}
\BIBentryALTinterwordspacing
Y.~Bengio, A.~Courville, and P.~Vincent, ``{Representation learning: A review
  and new perspectives},'' \emph{IEEE Transactions on Pattern Analysis and
  Machine Intelligence}, vol.~35, no.~8, pp. 1798--1828, 2013. [Online].
  Available: \url{https://arxiv.org/pdf/1206.5538.pdf}
\BIBentrySTDinterwordspacing

\bibitem{Litjens2017}
\BIBentryALTinterwordspacing
G.~Litjens, T.~Kooi, B.~E. Bejnordi, A.~A.~A. Setio, F.~Ciompi, M.~Ghafoorian,
  J.~A. W.~M. van~der Laak, B.~van Ginneken, and C.~I. S{\'{a}}nchez, ``{A
  survey on deep learning in medical image analysis.}'' \emph{Medical image
  analysis}, vol.~42, pp. 60--88, 2017. [Online]. Available:
  \url{http://www.ncbi.nlm.nih.gov/pubmed/28778026}
\BIBentrySTDinterwordspacing

\bibitem{He2015d}
\BIBentryALTinterwordspacing
K.~He, X.~Zhang, S.~Ren, and J.~Sun, ``{Delving Deep into Rectifiers:
  Surpassing Human-Level Performance on ImageNet Classification},''
  \emph{CoRR}, vol. abs/1502.0, 2015. [Online]. Available:
  \url{http://arxiv.org/abs/1502.01852}
\BIBentrySTDinterwordspacing

\bibitem{Lange1997}
\BIBentryALTinterwordspacing
N.~Lange, C.~M. Bishop, and B.~D. Ripley, ``{Neural Networks for Pattern
  Recognition.}'' \emph{Journal of the American Statistical Association},
  vol.~92, no. 440, p. 1642, dec 1997. [Online]. Available:
  \url{https://www.jstor.org/stable/2965437?origin=crossref}
\BIBentrySTDinterwordspacing

\bibitem{Bottou2010}
\BIBentryALTinterwordspacing
L.~Bottou, ``{Large-Scale Machine Learning with Stochastic Gradient Descent
  L´eon},'' \emph{Proceedings of COMPSTAT'2010}, 2010. [Online]. Available:
  \url{http://link.springer.com/10.1007/978-3-7908-2604-3}
\BIBentrySTDinterwordspacing

\bibitem{Kingma2014}
\BIBentryALTinterwordspacing
D.~P. Kingma and J.~Ba, ``{Adam: A Method for Stochastic Optimization},'' pp.
  1--15, 2014. [Online]. Available: \url{http://arxiv.org/abs/1412.6980}
\BIBentrySTDinterwordspacing

\bibitem{Osher2018}
\BIBentryALTinterwordspacing
S.~Osher, B.~Wang, P.~Yin, X.~Luo, M.~Pham, and A.~Lin, ``{Laplacian Smoothing
  Gradient Descent},'' pp. 1--28, 2018. [Online]. Available:
  \url{http://arxiv.org/abs/1806.06317}
\BIBentrySTDinterwordspacing

\bibitem{LeCun1998}
\BIBentryALTinterwordspacing
Y.~A. LeCun, L.~Bottou, G.~B. Orr, and K.-R. M{\"{u}}ller, ``{Efficient
  BackProp},'' \emph{Neural Networks, Tricks of the Trade}, 1998. [Online].
  Available: \url{http://yann.lecun.com/exdb/publis/pdf/lecun-98b.pdf
  http://link.springer.com/10.1007/978-3-642-35289-8{\_}3}
\BIBentrySTDinterwordspacing

\bibitem{Springenberg2014}
\BIBentryALTinterwordspacing
J.~T. Springenberg, A.~Dosovitskiy, T.~Brox, and M.~Riedmiller, ``{Striving for
  Simplicity: The All Convolutional Net},'' pp. 1--14, 2014. [Online].
  Available: \url{http://arxiv.org/abs/1412.6806}
\BIBentrySTDinterwordspacing

\bibitem{Long2015}
\BIBentryALTinterwordspacing
J.~Long, E.~Shelhamer, and T.~Darrell, ``{Fully convolutional networks for
  semantic segmentation},'' in \emph{2015 IEEE Conference on Computer Vision
  and Pattern Recognition (CVPR)}, vol. 2017-May.\hskip 1em plus 0.5em minus
  0.4em\relax IEEE, jun 2015, pp. 3431--3440. [Online]. Available:
  \url{https://people.eecs.berkeley.edu/{~}jonlong/long{\_}shelhamer{\_}fcn.pdf
  http://ieeexplore.ieee.org/document/7298965/}
\BIBentrySTDinterwordspacing

\bibitem{He2016}
\BIBentryALTinterwordspacing
K.~He, X.~Zhang, S.~Ren, and J.~Sun, ``{Deep Residual Learning for Image
  Recognition},'' in \emph{2016 IEEE Conference on Computer Vision and Pattern
  Recognition (CVPR)}, vol.~32, no.~5.\hskip 1em plus 0.5em minus 0.4em\relax
  IEEE, jun 2016, pp. 770--778. [Online]. Available:
  \url{http://joi.jlc.jst.go.jp/JST.JSTAGE/cl/2003.428?from=CrossRef
  http://ieeexplore.ieee.org/document/7780459/}
\BIBentrySTDinterwordspacing

\bibitem{Taha2015}
\BIBentryALTinterwordspacing
A.~A. Taha and A.~Hanbury, ``{Metrics for evaluating 3D medical image
  segmentation : analysis , selection , and tool},'' \emph{BMC Medical
  Imaging}, 2015. [Online]. Available:
  \url{http://dx.doi.org/10.1186/s12880-015-0068-x}
\BIBentrySTDinterwordspacing

\bibitem{Ronneberger2015}
O.~Ronneberger, P.~Fischer, and T.~Brox, ``{U-net: Convolutional networks for
  biomedical image segmentation},'' \emph{Lecture Notes in Computer Science
  (including subseries Lecture Notes in Artificial Intelligence and Lecture
  Notes in Bioinformatics)}, vol. 9351, pp. 234--241, 2015.

\bibitem{Drozdzal2016}
M.~Drozdzal, E.~Vorontsov, G.~Chartrand, S.~Kadoury, and C.~Pal, ``{The
  importance of skip connections in biomedical image segmentation},''
  \emph{Lecture Notes in Computer Science (including subseries Lecture Notes in
  Artificial Intelligence and Lecture Notes in Bioinformatics)}, vol. 10008
  LNCS, pp. 179--187, 2016.

\bibitem{Cicek2016}
{\"{O}}.~{\c{C}}i{\c{c}}ek, A.~Abdulkadir, S.~S. Lienkamp, T.~Brox, and
  O.~Ronneberger, ``{3D U-net: Learning dense volumetric segmentation from
  sparse annotation},'' \emph{Lecture Notes in Computer Science (including
  subseries Lecture Notes in Artificial Intelligence and Lecture Notes in
  Bioinformatics)}, vol. 9901 LNCS, pp. 424--432, 2016.

\bibitem{Milletari2016}
F.~Milletari, N.~Navab, and S.~A. Ahmadi, ``{V-Net: Fully convolutional neural
  networks for volumetric medical image segmentation},'' \emph{Proceedings -
  2016 4th International Conference on 3D Vision, 3DV 2016}, pp. 565--571,
  2016.

\bibitem{Myronenko2018}
\BIBentryALTinterwordspacing
A.~Myronenko, ``{3D MRI brain tumor segmentation using autoencoder
  regularization},'' 2018. [Online]. Available:
  \url{http://arxiv.org/abs/1810.11654}
\BIBentrySTDinterwordspacing

\bibitem{Kingma2013}
\BIBentryALTinterwordspacing
D.~P. Kingma and M.~Welling, ``{Auto-Encoding Variational Bayes},'' in
  \emph{Proceedings of the 2nd International Conference on Learning
  Representations (ICLR)}, no. 2014, dec 2013, pp. 1--14. [Online]. Available:
  \url{http://arxiv.org/abs/1312.6114}
\BIBentrySTDinterwordspacing

\bibitem{Wu2018}
\BIBentryALTinterwordspacing
Y.~Wu and K.~He, ``{Group normalization},'' \emph{Lecture Notes in Computer
  Science (including subseries Lecture Notes in Artificial Intelligence and
  Lecture Notes in Bioinformatics)}, vol. 11217 LNCS, pp. 3--19, 2018.
  [Online]. Available: \url{https://arxiv.org/pdf/1803.08494.pdf}
\BIBentrySTDinterwordspacing

\bibitem{Kamnitsas2018}
\BIBentryALTinterwordspacing
K.~Kamnitsas, W.~Bai, E.~Ferrante, S.~McDonagh, M.~Sinclair, N.~Pawlowski,
  M.~Rajchl, M.~Lee, B.~Kainz, D.~Rueckert, and B.~Glocker, ``{Ensembles of
  multiple models and architectures for robust brain tumour segmentation},''
  \emph{Lecture Notes in Computer Science (including subseries Lecture Notes in
  Artificial Intelligence and Lecture Notes in Bioinformatics)}, vol. 10670
  LNCS, pp. 450--462, 2018. [Online]. Available:
  \url{https://arxiv.org/pdf/1711.01468.pdf}
\BIBentrySTDinterwordspacing

\bibitem{Isensee2018}
\BIBentryALTinterwordspacing
F.~Isensee, P.~Kickingereder, W.~Wick, M.~Bendszus, and K.~H. Maier-Hein, ``{No
  New-Net},'' 2018. [Online]. Available: \url{http://arxiv.org/abs/1809.10483}
\BIBentrySTDinterwordspacing

\bibitem{Li2018}
\BIBentryALTinterwordspacing
H.~Li, J.~Zhang, M.~Muehlau, J.~Kirschke, and B.~Menze, ``{Multi-Scale
  Convolutional-Stack Aggregation for Robust White Matter Hyperintensities
  Segmentation},'' 2018. [Online]. Available:
  \url{http://arxiv.org/abs/1807.05153}
\BIBentrySTDinterwordspacing

\bibitem{plummer1998emergency}
D.~Plummer, J.~Clinton, and B.~Matthew, ``Emergency department ultrasound
  improves time to diagnosis and survival in ruptured abdominal aortic
  aneurysm,'' \emph{Acad Emerg Med}, vol.~5, no.~5, p. 417, 1998.

\bibitem{Rodgerson2001}
J.~D. Rodgerson, W.~G. Heegaard, D.~Plummer, J.~Hicks, J.~Clinton, and
  S.~Sterner, ``{Emergency department right upper quadrant ultrasound is
  associated with a reduced time to diagnosis and treatment of ruptured ectopic
  pregnancies},'' \emph{Academic Emergency Medicine}, vol.~8, no.~4, pp.
  331--336, 2001.

\bibitem{Barnes2005}
\BIBentryALTinterwordspacing
T.~W. Barnes, T.~I. Morgenthaler, E.~J. Olson, G.~K. Hesley, P.~A. Decker, and
  J.~H. Ryu, ``{Sonographically guided thoracentesis and rate of
  pneumothorax},'' \emph{Journal of Clinical Ultrasound}, vol.~33, no.~9, pp.
  442--446, nov 2005. [Online]. Available:
  \url{http://doi.wiley.com/10.1002/jcu.20163}
\BIBentrySTDinterwordspacing

\bibitem{Bauman2009}
M.~Bauman, D.~Braude, and C.~Crandall, ``{Ultrasound-guidance vs. standard
  technique in difficult vascular access patients by ED technicians},''
  \emph{American Journal of Emergency Medicine}, vol.~27, no.~2, pp. 135--140,
  2009.

\bibitem{Barnett2018}
\BIBentryALTinterwordspacing
M.~Barnett, M.~T. Pillow, J.~Carnell, A.~Rohra, S.~DeSandro, and A.~K. Gardner,
  ``{Informing the Revolution: A Needs Assessment of Ultrasound Knowledge and
  Skills Among Graduating Physician Assistant Students},'' \emph{The Journal of
  Physician Assistant Education}, p.~1, aug 2018. [Online]. Available:
  \url{http://insights.ovid.com/crossref?an=01367895-900000000-99916}
\BIBentrySTDinterwordspacing

\bibitem{Lewiss2014}
R.~E. Lewiss, B.~Hoffmann, Y.~Beaulieu, and M.~B. Phelan, ``{Point-of-Care
  ultrasound education},'' \emph{Journal of Ultrasound in Medicine}, vol.~33,
  no.~1, pp. 27--32, 2014.

\bibitem{Truong2016}
\BIBentryALTinterwordspacing
K.~Truong, A.~Janssen, C.~Moore, and V.~LaBond, ``{326 Use of a Didactic Low
  Fidelity Simulated Model to Measure Objective Improvement in Corneal Foreign
  Body Removal},'' \emph{Annals of Emergency Medicine}, vol.~68, no.~4, pp.
  S125--S126, oct 2016. [Online]. Available:
  \url{http://dx.doi.org/10.1016/j.annemergmed.2016.08.342
  https://linkinghub.elsevier.com/retrieve/pii/S0196064416308009}
\BIBentrySTDinterwordspacing

\bibitem{Rowley2019}
\BIBentryALTinterwordspacing
K.~Rowley, K.~Wheeler, D.~Pruthi, D.~Kaushik, A.~Mansour, J.~Basler, and
  M.~Liss, ``{MP80-19 DEVELOPMENT AND IMPLEMENTATION OF COMPETENCY-BASED
  ASSESSMENT FOR UROLOGICAL ULTRASOUND TRAINING USING HUMAN MODELS AND SONOSIM
  TESTING},'' \emph{Journal of Urology}, vol. 201, no. Supplement 4, pp.
  38--39, apr 2019. [Online]. Available:
  \url{http://www.jurology.com/doi/10.1097/01.JU.0000557418.26841.d9}
\BIBentrySTDinterwordspacing

\bibitem{Silva2016}
\BIBentryALTinterwordspacing
J.~P. Silva, T.~Plescia, N.~Molina, A.~{Claudia De Oliveira Tonelli},
  M.~Langdorf, and J.~C. Fox, ``{Journal of Educational Evaluation for Health
  Professions Randomized study of effectiveness of computerized ultrasound
  simulators for an introductory course for residents in Brazil},'' \emph{J
  Educ Eval Health Prof}, vol.~13, 2016. [Online]. Available:
  \url{http://dx.doi.org/10.3352/jeehp.2016.13.16}
\BIBentrySTDinterwordspacing

\bibitem{Chung2013}
\BIBentryALTinterwordspacing
G.~K. W.~K. Chung, R.~G. Gyllenhammer, E.~L. Baker, and E.~Savitsky, ``{Effects
  of Simulation-Based Practice on Focused Assessment With Sonography for Trauma
  (FAST) Window Identification, Acquisition, and Diagnosis},'' \emph{Military
  Medicine}, vol. 178, no. 10S, pp. 87--97, 2013. [Online]. Available:
  \url{https://watermark.silverchair.com/milmed-d-13-00208.pdf?}
\BIBentrySTDinterwordspacing

\bibitem{Paddock2015}
M.~Paddock, J.~Bailitz, R.~Horowitz, B.~Khishfe, K.~Cosby, and M.~Sergel,
  ``{Disaster Response Team FAST Skills Training with a PortableUltrasound
  Simulator Compared to Traditional Training: Pilot Study},'' \emph{Western
  Journal of Emergency Medicine}, vol.~16, no.~2, pp. 325--330, 2015.

\bibitem{Zosa2018}
T.~E. Zosa, M.~Wang, and E.~Savitsky, ``{Deep Learning for Ultrasound Image
  Enhancement},'' in \emph{CRESST Conference 2018}, Los Angeles, 2018.

\bibitem{Nandamuri2019}
\BIBentryALTinterwordspacing
S.~Nandamuri, D.~China, P.~Mitra, and D.~Sheet, ``{SUMNet: Fully Convolutional
  Model for Fast Segmentation of Anatomical Structures in Ultrasound
  Volumes},'' pp. 6--9, 2019. [Online]. Available:
  \url{http://arxiv.org/abs/1901.06920}
\BIBentrySTDinterwordspacing

\bibitem{Ghesu2016}
F.~C. Ghesu, E.~Krubasik, B.~Georgescu, V.~Singh, Y.~Zheng, J.~Hornegger, and
  D.~Comaniciu, ``{Marginal Space Deep Learning: Efficient Architecture for
  Volumetric Image Parsing},'' \emph{IEEE Transactions on Medical Imaging},
  vol.~35, no.~5, pp. 1217--1228, 2016.

\bibitem{Dong2018}
S.~Dong, G.~Luo, K.~Wang, S.~Cao, Q.~Li, and H.~Zhang, ``{A Combined Fully
  Convolutional Networks and Deformable Model for Automatic Left Ventricle
  Segmentation Based on 3D Echocardiography},'' \emph{BioMed Research
  International}, vol. 2018, pp. 1--16, 2018.

\bibitem{Bernard2016}
O.~Bernard, J.~G. Bosch, B.~Heyde, M.~Alessandrini, D.~Barbosa,
  S.~Camarasu-Pop, F.~Cervenansky, S.~Valette, O.~Mirea, M.~Bernier, P.~M.
  Jodoin, J.~S. Domingos, R.~V. Stebbing, K.~Keraudren, O.~Oktay, J.~Caballero,
  W.~Shi, D.~Rueckert, F.~Milletari, S.~A. Ahmadi, E.~Smistad, F.~Lindseth,
  M.~{Van Stralen}, C.~Wang, {\"{O}}.~Smedby, E.~Donal, M.~Monaghan,
  A.~Papachristidis, M.~L. Geleijnse, E.~Galli, and J.~D'Hooge, ``{Standardized
  Evaluation System for Left Ventricular Segmentation Algorithms in 3D
  Echocardiography},'' \emph{IEEE Transactions on Medical Imaging}, vol.~35,
  no.~4, pp. 967--977, 2016.

\bibitem{Dong2018a}
S.~Dong, G.~Luo, K.~Wang, S.~Cao, A.~Mercado, O.~Shmuilovich, H.~Zhang, and
  S.~Li, ``{VoxelAtlasGAN: 3D Left Ventricle Segmentation on Echocardiography
  with Atlas Guided Generation and Voxel-to-Voxel Discrimination},''
  \emph{Lecture Notes in Computer Science (including subseries Lecture Notes in
  Artificial Intelligence and Lecture Notes in Bioinformatics)}, vol. 11073
  LNCS, pp. 622--629, 2018.

\bibitem{Badrinarayanan2017}
\BIBentryALTinterwordspacing
V.~Badrinarayanan, A.~Kendall, and R.~Cipolla, ``{SegNet: A Deep Convolutional
  Encoder-Decoder Architecture for Image Segmentation},'' \emph{IEEE
  Transactions on Pattern Analysis and Machine Intelligence}, vol.~39, no.~12,
  pp. 2481--2495, dec 2017. [Online]. Available:
  \url{https://ieeexplore.ieee.org/document/7803544/}
\BIBentrySTDinterwordspacing

\bibitem{Wunderling2017}
T.~Wunderling, B.~Golla, P.~Poudel, C.~Arens, M.~Friebe, and C.~Hansen,
  ``{Comparison of thyroid segmentation techniques for 3D ultrasound},'' in
  \emph{SPIE Medical Imaging 2017}, M.~A. Styner and E.~D. Angelini, Eds., vol.
  10133, feb 2017.

\bibitem{Dollar}
\BIBentryALTinterwordspacing
P.~Dollar, Z.~Tu, P.~Perona, and S.~Belongie, ``{Integral Channel Features},''
  in \emph{Procedings of the British Machine Vision Conference 2009}.\hskip 1em
  plus 0.5em minus 0.4em\relax British Machine Vision Association, 2009, pp.
  91.1--91.11. [Online]. Available:
  \url{http://www.bmva.org/bmvc/2009/Papers/Paper244/Paper244.html}
\BIBentrySTDinterwordspacing

\bibitem{Xie2017}
\BIBentryALTinterwordspacing
S.~Xie, R.~Girshick, P.~Dollar, Z.~Tu, and K.~He, ``{Aggregated Residual
  Transformations for Deep Neural Networks},'' in \emph{2017 IEEE Conference on
  Computer Vision and Pattern Recognition (CVPR)}.\hskip 1em plus 0.5em minus
  0.4em\relax IEEE, jul 2017, pp. 5987--5995. [Online]. Available:
  \url{http://arxiv.org/abs/1611.05431v2
  http://ieeexplore.ieee.org/document/8100117/}
\BIBentrySTDinterwordspacing

\bibitem{Zhang2018}
\BIBentryALTinterwordspacing
Y.~Zhang, K.~Li, K.~Li, L.~Wang, B.~Zhong, and Y.~Fu, ``{Image Super-Resolution
  Using Very Deep Residual Channel Attention Networks},'' in \emph{15th
  European Conference on Computer Vision (ECCV)}, jul 2018, pp. 294--310.
  [Online]. Available: \url{http://arxiv.org/abs/1807.02758
  http://link.springer.com/10.1007/978-3-030-01234-2{\_}18}
\BIBentrySTDinterwordspacing

\bibitem{Xie2018}
S.~Xie, C.~Sun, J.~Huang, Z.~Tu, and K.~Murphy, ``{Rethinking Spatiotemporal
  Feature Learning: Speed-Accuracy Trade-offs in Video Classification},'' in
  \emph{15th European Conference on Computer Vision (ECCV)}, no.~1, 2018, pp.
  318--335.

\bibitem{Shan2018}
\BIBentryALTinterwordspacing
H.~Shan, Y.~Zhang, Q.~Yang, U.~Kruger, M.~K. Kalra, L.~Sun, W.~Cong, and
  G.~Wang, ``{3-D Convolutional Encoder-Decoder Network for Low-Dose CT via
  Transfer Learning From a 2-D Trained Network},'' \emph{IEEE Transactions on
  Medical Imaging}, vol.~37, no.~6, pp. 1522--1534, jun 2018. [Online].
  Available: \url{https://ieeexplore.ieee.org/document/8353466/}
\BIBentrySTDinterwordspacing

\end{thebibliography}

\end{document}